\documentclass[journal]{IEEEtran}
\usepackage{xcolor,soul,framed}
\usepackage{graphicx}  % Written by David Carlisle and Sebastian Rahtz
\usepackage{epstopdf}

\usepackage{nomencl}
\makenomenclature

\makeatletter
\newcommand{\removelatexerror}{\let\@latex@error\@gobble}
\makeatother
\let\chapter\section
\usepackage{algorithm}
\usepackage{algpseudocode}
\usepackage{algorithmicx}

\usepackage{array}
\usepackage{arydshln}
\usepackage{blkarray}

\usepackage{pifont}

\usepackage{multirow}
\usepackage{booktabs}
\usepackage{subfloat}
\usepackage{subfigure}
\usepackage{url}
\usepackage{stfloats}
\usepackage{color}

\usepackage{amsmath}
\usepackage{amsfonts}
\usepackage{amssymb}
\usepackage{amsthm}

\DeclareMathOperator*{\argmin}{arg\,min}

\renewcommand{\baselinestretch}{1}

\renewcommand{\baselinestretch}{1}

\ifCLASSINFOpdf

\else

\fi

\usepackage{etoolbox}
\renewcommand
\nomgroup[1]{%
\item[\bfseries
\ifstrequal{#1}{A}{{Abbreviations}}%
]
\item[\bfseries
\ifstrequal{#1}{B}{{Parameters and Constants}}%
]
\item[\bfseries
\ifstrequal{#1}{C}{{Variables}}%
]}
\usepackage{tabularx}
\usepackage{setspace}
\usepackage{threeparttable}
\usepackage{bm}

\begin{document}
%
% paper title
% Titles are generally capitalized except for words such as a, an, and, as,
% at, but, by, for, in, nor, of, on, or, the, to and up, which are usually
% not capitalized unless they are the first or last word of the title.
% Linebreaks \\ can be used within to get better formatting as desired.
% Do not put math or special symbols in the title.

%\title{A Review of Vulnerability Assessment for the AI-enabled Power System Applications}

\title{Vulnerability of Machine Learning Approaches Applied in IoT-based Smart Grid:  A Review}

\author{
Zhenyong Zhang, Mengxiang Liu, \IEEEmembership{Member,~IEEE}, \\ Mingyang Sun, \IEEEmembership{Senior member,~IEEE}, Ruilong Deng, \IEEEmembership{Senior  member,~IEEE}, Peng Cheng, \IEEEmembership{Member,~IEEE},  \\  Dusit Niyato, \IEEEmembership{Fellow,~IEEE},  Mo-Yuen Chow, \IEEEmembership{Fellow,~IEEE}, and Jiming Chen, \IEEEmembership{Fellow,~IEEE}
%Ruilong Deng,~\IEEEmembership{Senior Member,~IEEE,}
%Rongxing Lu,~\IEEEmembership{Fellow,~IEEE}
\thanks{Z. Zhang is with the State Key Laboratory of Public Big Data, College of Computer Science and Technology, Guizhou University, Guiyang 550000, China, e-mail: zyzhangnew@gmail.com;

M. Liu is with the Department of Automatic Control and Systems Engineering, University of Sheffield, Sheffield S10 2TN, UK, e-mail: liumengxiang329@gmail.com;

M. Sun, R. Deng, P. Cheng, and J. Chen are with the State Key Laboratory of Industrial Control Technology and the College of Control Science and Engineering, Zhejiang University, Hangzhou 310027, China, e-mail:  mingyangsun@zju.edu.cn, dengruilong@zju.edu.cn, lunarheart@zju.edu.cn, cjm@zju.edu.cn;

D. Niyato is with the School of Computer Science and Engineering, Nanyang Technological University, Singapore 639798, e-mail: dniyato@ntu.edu.sg;

M.-Y. Chow is with the UM–SJTU Joint Institute Shanghai Jiao Tong University, Shanghai 200240, China, e-mail: moyuen.chow@sjtu.edu.cn.
}

}
%\thanks{The authors are with the State Key Laboratory of Industrial Control Technology and the College of Control Science and Engineering, Zhejiang University, Hangzhou 310027, China;}}
\renewcommand{\baselinestretch}{1}
\markboth{Prepared for IEEE IoT-J}%
{Shell \MakeLowercase{\textit{et al.}}: Bare Demo of IEEEtran.cls for IEEE Journals}

% make the title area
\maketitle

% As a general rule, do not put math, special symbols or citations
% in the abstract
\begin{abstract}
Machine learning (ML) sees an increasing prevalence of being used in the internet-of-things (IoT)-based smart grid. However, the trustworthiness of ML is a severe issue that must be addressed to accommodate the trend of ML-based smart grid applications (MLsgAPPs). The adversarial distortion injected into the power signal will greatly affect the system's normal control and operation. Therefore, it is imperative to conduct vulnerability assessment for MLsgAPPs applied in the context of safety-critical power systems. In this paper, we provide a comprehensive review of the recent progress in designing attack and defense methods for MLsgAPPs. Unlike the traditional survey about ML security, this is the first review work about the security of MLsgAPPs that focuses on the characteristics of power systems. We first highlight the specifics for constructing the adversarial attacks on MLsgAPPs. Then, the vulnerability of MLsgAPP is analyzed from both the aspects of the power system and ML model. Afterward, a comprehensive survey is conducted to review and compare existing studies about the adversarial attacks on MLsgAPPs in scenarios of generation, transmission, distribution, and consumption, and the countermeasures are reviewed according to the attacks that they defend against. Finally, the future research directions are discussed on the attacker's and defender's side, respectively. We also analyze the potential vulnerability of large language model-based (e.g., ChatGPT) power system applications. Overall, we encourage more researchers to contribute to investigating the adversarial issues of MLsgAPPs.
\end{abstract}

\IEEEpeerreviewmaketitle

\begin{IEEEkeywords}
Smart Grid, Vulnerability Assessment, Adversarial Machine Learning, Power System Specifics, Physical constraints \end{IEEEkeywords}

\mbox{}
\nomenclature[A, 01]{\textbf{IoT}}{Internet-of-Things}
\nomenclature[A, 02]{\textbf{ML}}{Machine Learning}
\nomenclature[A, 03]{\textbf{PMU}}{Phasor Measurement Unit}
\nomenclature[A, 04]{\textbf{SCADA}}{Supervisory Control And Data Acquisition}
\nomenclature[A, 05]{\textbf{AI}}{Artificial Intelligence}
\nomenclature[A, 06]{\textbf{SVM}}{Support Vector Machine}
\nomenclature[A, 07]{\textbf{DL}}{Deep Learning}
\nomenclature[A, 07]{\textbf{LR}}{Logistic Regression}
\nomenclature[A, 08]{\textbf{DT}}{Decision Tree}
\nomenclature[A, 09]{\textbf{MLP}}{Multi-Layer Perception}
\nomenclature[A, 10]{\textbf{RES}}{Renewable Energy Source}
\nomenclature[A, 11]{\textbf{RNN}}{Recurrent Neural Network}
\nomenclature[A, 12]{\textbf{CNN}}{Convolutional Neural Network}
\nomenclature[A, 13]{\textbf{DRL}}{Deep Reinforcement Learning}
\nomenclature[A, 14]{\textbf{MLsgAPP}}{Machine Learning-based smart grid APPlication}
\nomenclature[A, 15]{\textbf{CV}}{Computer Vision}
\nomenclature[A, 16]{\textbf{CPS}}{Cyber-Physical System}
\nomenclature[A, 17]{\textbf{NN}}{Neural Network}
\nomenclature[A, 18]{\textbf{DNN}}{Deep Neural Network}
\nomenclature[A, 19]{\textbf{SA}}{Scaling Attack}
\nomenclature[A, 20]{\textbf{FGSM}}{Fast Gradient Sign Method}
\nomenclature[A, 21]{\textbf{FGV}}{Fast Gradient Value}
\nomenclature[A, 22]{\textbf{PGD}}{Projected Gradient Descent}
\nomenclature[A, 23]{\textbf{JSMA}}{Jacobian-based Salience Map Attack}
\nomenclature[A, 24]{\textbf{BIM}}{Basic Iterative Method}
\nomenclature[A, 25]{\textbf{SE}}{State Estimation}
\nomenclature[A, 26]{\textbf{BDD}}{Bad Data Detection}
\nomenclature[A, 27]{\textbf{UVLS}}{Under Voltage Load Shedding}
\nomenclature[A, 28]{\textbf{AMI}}{Advanced Metering Infrastructure}
\nomenclature[A, 29]{\textbf{NESCOR}}{National Electric Sector Cybersecurity Organization Resource}
\nomenclature[A, 30]{\textbf{WAMPAC}}{Wide-Area Monitoring, Protection, and Control}
\nomenclature[A, 31]{\textbf{FDIA}}{False Data Injection Attack}
\nomenclature[A, 32]{\textbf{GAN}}{Generative Adversarial Network}
\nomenclature[A, 33]{\textbf{ELM}}{Extreme Learning Machine}
\nomenclature[A, 34]{\textbf{CCT}}{Critical Clearing Time}
\nomenclature[A, 35]{\textbf{AC}}{Alternating Current}
\nomenclature[A, 36]{\textbf{DC}}{Direct Current}
\nomenclature[A, 37]{\textbf{RII}}{Robustness Index for Instance}
\nomenclature[A, 38]{\textbf{RIC}}{Robustness Index for Classifier}
\nomenclature[A, 39]{\textbf{MAPE}}{Maximum Absolute Percentage Error}
\nomenclature[A, 40]{\textbf{BNN}}{Belief Neural Network}
\nomenclature[A, 41]{\textbf{EPD}}{Expected Performance Decay}
\nomenclature[A, 42]{\textbf{EPDR}}{Expected Performance Decay Rate}
\nomenclature[A, 43]{\textbf{DBN}}{Deep Belief Network}
\nomenclature[A, 44]{\textbf{MILP}}{Mixed Integer Linear Programming}
\nomenclature[A, 45]{\textbf{ReLU}}{Rectified Linear Unit}
\nomenclature[A, 46]{\textbf{ReLU}}{Short-Term Voltage Stability}
\nomenclature[A, 47]{\textbf{LSTM}}{Long Short-Term Memory}
\nomenclature[A, 48]{\textbf{FCNN}}{Fully Convolutional Neural Network}
\nomenclature[A, 49]{\textbf{BPNN}}{Back-Propagation Neural Network}
\nomenclature[A, 50]{\textbf{ELU}}{Exponential Linear Units}
\nomenclature[A, 51]{\textbf{DR}}{Demand Response}
\nomenclature[A, 52]{\textbf{MLR}}{Multiple Linear Regressor}
\nomenclature[A, 53]{\textbf{API}}{Application Programming Interface}
\nomenclature[A, 54]{\textbf{NILM}}{Non-Intrusive Load Monitoring}
\nomenclature[A, 55]{\textbf{ACOPF}}{AC Optimal Power Flow}
\nomenclature[A, 56]{\textbf{SCOPF}}{Security-Constrained Optimal Power Flow}
\nomenclature[A, 57]{\textbf{MARL}}{Multi-Agent Reinforcement Learning}
\nomenclature[A, 58]{\textbf{NYISO}}{New York Independent
System Operator}
\nomenclature[A, 59]{\textbf{L-BFGS}}{Limited-memory Broyden-Fletcher-Goldfarb-Shanno}
\nomenclature[A, 60]{\textbf{FNN}}{Forward Neural Network}
\nomenclature[A, 61]{\textbf{ARIMA}}{Autoregressive Integrated Moving Average Model}
\nomenclature[A, 62]{\textbf{FFNN}}{Feed-forward Neural Network}
\nomenclature[A, 63]{\textbf{RTDS}}{Real Time Digital Simulator}
\nomenclature[A, 64]{\textbf{RLGC}}{Reinforcement Learning for Grid Control}
\nomenclature[A, 65]{\textbf{LLM}}{Large Language Model}
\printnomenclature

\section{Introduction}\label{section:introduction}
Decarbonization, digitization, and decentralization are key drivers to achieving the ambitious goal of carbon neutrality all over the world. The high-end electronic and communication infrastructures result in a huge reduction of the dependency on fossil fuels in power systems. However, although the IoT technologies facilitate the energy transition from low-efficient to smart and flexible operation, they also bring in challenges such as massive operational uncertainties, lower system inertia, complex system dynamics, and highly nonlinear models, which make it difficult to derive accurate mathematical models as well as efficient operation and control solutions. Therefore, it is imperative to take advantage of the cutting-edge ML approaches.

ML is likely to be widely used in power systems as plenty of supporting infrastructures have been deployed. With the advent of advanced sensors [e.g., smart meters and PMUs], it has the potential to collect fine-grained electricity data from field devices. Specifically, the SCADA system plays an important role in connecting hundreds of thousands of substations and millions of communication devices to gather different types of data (discrete/continuous, digital/analog, structured/unstructured), such as maintenance logs, breaker/switch's status, current and voltage measurements, and the data from the external system such as the weather forecast, wind speed, and solar magnetic. Besides, by adopting cloud and edge computing, the computation power of the power system is largely improved \cite{li2022eccsg}. The combination of advanced data-collection approaches and cloud computing has enabled the use of complex ML models due to the high-performance computing, scalable structure, and prodigious storage capability, they afford \cite{konstantelos2016parallel}.

%Furthermore, the computation power of power system is enhanced with the introduction of networked computer servers, supercomputers, and fast data-transmission links. Equipped with controllable power devices such as synchronous generators, microgrid converters, high/medium-voltage direct current (H/MVDC), and flexible AC transmission system (FACTS), the power system becomes a highly automated system and has the capability of preventing outages. The digitalized control process makes it faster to response to the decisions made by the control center. The feedback loop is automatically closed by integrating the communication, computing, and control infrastructures, which makes it promising to achieve real-time operation and control for the large-scale power system using ML approaches.

%Actually, the modern power system is the largest and smartest networked physical system that human has ever built.
Since modern society highly depends on electricity, every country spends a lot of financial and research resources on developing and running the power system efficiently and safely. With the coming of the AI era, many countries all over the world are moving toward operating their grids using ML. In 2020, the U.S. Department of Energy put forward that the use of AI technologies can make the computation and operation converge and enable the automatic control of required functions.\footnote{2020 Smart Grid System Report. Office of Electricity. https://www.energy.gov/oe/articles/2020-smart-grid-system-report} Europe also attempts to use AI technologies to support power market activities \cite{niet2021europeact} and reduce carbon emission\footnote{Digitalisation of the European Energy System. European Commission. https://digital-strategy.ec.europa.eu/en/policies/digitalisation-energy}. The Chinese government launched the 13th plan for the development of power systems by encouraging the implementation of AI technologies in cyber systems\footnote{http://www.gov.cn/xinwen/} and the 14th plan to further improve the grid's intelligent level of control and operations. Japan \cite{japanai2022} and Korea\footnote{Data and Artificial Intelligence Economic Activation Plan (2019-2023). Ministry of Science and ICT. https://eiec.kdi.re.kr/policy/materialView.do?num=184568} also show their resolutions to promote AI applied in power systems, catching up with the worldwide race in this track.

Therefore, the application of ML in power systems is attracting increasing attention from both academia and industry. Since 1970s, plenty of studies have proposed the implementation framework for using the ML approaches in power system applications such as dynamic security assessment \cite{doaiswami1979ifac,sobajic1989anndsa}, wind/solar power and building load forecasting \cite{chen2018dlunsupe, hong2016gefcompete}, power outage detection \cite{eskandarpour2017podetec}, user's profile classification \cite{yang2018userprofile}, heating, ventilation, and air conditioning\ control and grid protection \cite{lassetter2018havccontrol}. Before Hinton's first proposal about DL \cite{hiton2006dl}, researchers have tried their attempts to deal with complicated classification and regression tasks using shallow learning approaches such as SVM, LR, DT, and MLP. A typical application of these approaches is to assist the dynamic security assessment to reduce the computation burdens of executing time-domain simulations \cite{gomez2011svpdsa}.

Today, DL approaches are introduced to address the coordination and environmental challenges brought in by RES, microgrids, and power electronics. The RNN is widely used for load forecast as it has good performance in characterizing the temporal dependencies between feature inputs and predicted values \cite{kong2017dnnforecastload}. The critical data processing function, state estimation, is solved using the physics-guided deep autoencoder to deal with the nonlinear relationship between variables and real-time and convergence requirements in large-scale power systems \cite{wang2020aese}. To extract the spatial correlation, the faults are detected and located using the CNN in the distribution network with distributed energy resources \cite{rai2021faulclassi}. The DRL approaches are used for short-term emergency control, Volt/VAr control, long-term residential demand response, and battery energy management \cite{zhang2018drps}. The MLsgAPPs can achieve state-of-the-art performance compared to traditional model-based approaches and have real-time and complexity-resolved properties.

The ML algorithm can be embedded into the power system's running software to make decisions. For example, the iTesla project develops a dynamic security assessment toolbox based on DT to support the stable decision-making for the pan-European electricity transmission system \cite{vascondelos2016iteslaproj}. The DT-based security assessment is also applied to and validated with the Entergy system in New Orleans \cite{liu2014dsadtdt}. The French transmission system operator RTE aims to exploit the potential of ML to run the power network safely and autonomously \cite{lerps2020yoon}. Overall, the real-time requirement and modeling difficulty of system behaviors are two main promotional factors that foster the use of ML in power systems.

\begin{figure}[htbp!]
\begin{center}
\includegraphics[width=0.49\textwidth]{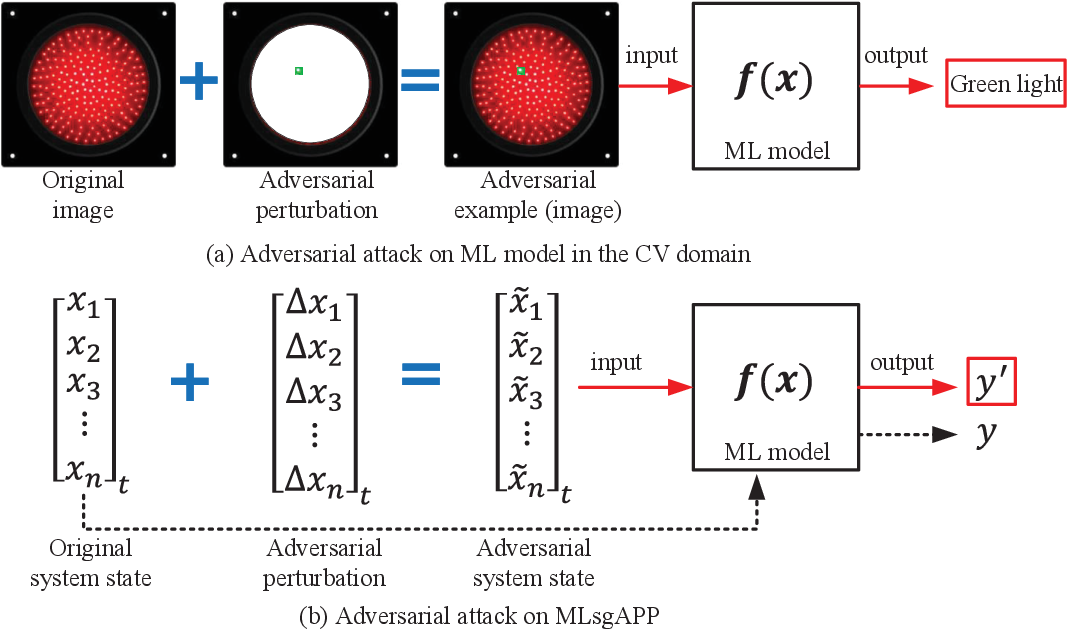}
\caption{The adversarial attack on the ML model and MLsgAPP model, respectively.}\label{fig:adversarialexample}
\end{center}
\vspace{-0.5cm}
\end{figure}

However, the malfunction of ML models raises a lot of concerns for their real-world implementation. The very early work \cite{dalvi2004emailfiltering} shows that the linear classifier can be cheated by spam emails with carefully crafted changes in text. The detection accuracy of web spam drops 50\% in the worst case if the SVM's input is added controlled perturbation \cite{zhou2012kdd}. The backdoor inserted into the training dataset manipulates the output of some specific inputs matching the backdoor trick \cite{gu2017backdoor}. Dated back to the 1980s, many pioneer researches have shown that the malfunction of ML could be exploited by adding noises or small perturbations to inputs \cite{kearns1988malicious,lowd1988malicious}. Although DL can achieve much more accurate performance (say nearly 100\%), the panda can still be misclassified as a giraffe by changing the input picture with very few pixels \cite{Szegedy2013rss}. The well-crafted malicious input replacing the clean data is termed as an adversarial example (as shown in Fig. \ref{fig:adversarialexample}), which is designed to be hardly perceptible to human eyes. %Since then, the adversarial ML becomes a growing concern in the computer vision (CV) society.
Although the development of computing structures such as the graphics processing unit creates a gloomy image processing industry, the deterioration of the performance of ML is inevitable due to the presence of adversarial examples \cite{opp2021aisecurity}. Adversarial machine learning poses an evident threat to ML-based applications in different areas, especially safety-critical systems such as the power system \cite{olowononi2021aisecuritycps}.

%The adversarial example is a threat since the ML model lacks the experience to make decisions with adversely perturbed input. A natural guess is that the DRL approaches also face the threat of adversarial examples. The truth is that the four critical components such as environment, observation, policy, and reward of the DRL method are all vulnerable to data integrity attacks \cite{Ilahi2021drlvul}. Huang \emph{et al.} \cite{huang2017drlvulne} also proved this point by validating the feasibility of the existing adversarial attack to undermine the DRL model used in different scenarios such as Atari games, autonomous driving, robot control tasks, and power system control. The Autopilot system of Tesla sometimes misrecognize the light and causes serious traffic accident\footnote{https://techcrunch.com/2022/05/18/nhtsa-probes-tesla-autopilot-crash-that-killed-three-people/}. Every day, there are millions of spam emails that bypass the detection system reaching the users' mails \cite{palka2015spamemail}. Fake videos and images can make the fake into truth without being identified\footnote{https://deepfakesweb.com/}. People are confused by what they see and operate.

What is worse is that cyberattacks can be easily introduced into the power system due to the proliferation of high-end sensing and control devices, creating a negative impact on the system's normal operation and control, which makes the situation worse when the defect of the ML model is not sufficiently addressed.
%The large-scale deployment of IoT devices encourages the use of marginally cheaper hardware which results in limited support of useful security measures. The AI-enabled applications highly rely on the synchronous data collected by the PMU network and slowly changed power data collected by the SCADA system. However, it has been widely conceived that communication networks (e.g., PMU and SCADA networks) are prone to attacks from malicious cyber actors \cite{yan2012sgcsecur}. It was first warned that the collected data by the SCADA system could be stealthily manipulated \cite{liu2011false}. The PMUs also lack of well protection and can be compromised remotely \cite{zhang2017fdipmu}.
A lot of studies have demonstrated the devastating effect caused by cyberattacks on power systems \cite{liang2017fdais}. Apart from the academic concern, the North American Electric Reliability Corporation also seriously considers the smart grid cybersecurity issue \cite{nerc2021security}. By building a botnet of computers, the European grid could lose synchronization if more than 2.5 million devices are infected by power consumption modification malware \cite{dabrowski2017gridshock}. The attack events that happened in Ukraine and Iran show that the threat of security breaches is of real existence \cite{case2016analysis,iran2010iran}. The authenticity of data steams is of great importance for downstream power system applications. The safety-critical power system might suffer devastating consequences subject to adversarial attacks on the operation data.
%Due to the strict requirement for data quality, high-end information and communication systems are deployed, but meanwhile it also introduces cyberattacks such as data integrity attacks.

The above facts show that the attacker has the capability of hacking into the cyberspace of power systems. The function depending on the information and communication infrastructure might face a great risk of being affected. Considering the data integrity attack, the input of ML cannot be assumed intact anymore. \textbf{Nevertheless, the adversarial issue of MLsgAPP is currently overlooked by researchers.}
%The related proposals are mainly about the adoption of existing attack approaches designed for image-related adversarial examples.
If the prediction of MLsgAPP is corrupted, not only could there be equipment failures or financial losses, but also in extremely realistic cases, the bulk power system assets could be inflicted, resulting in safety hazards. Chen \emph{et al.} \cite{chen2022fistml} put forward the first question about the vulnerability of the NN-based power quality classification and building energy consumption considering the threat of adversarial examples. The inherent vulnerability of NN makes NN-based applications prone to adversarial attacks. Therefore, there exists a philosophical dilemma: \textbf{the applied ML technique is like a double-edged sword, and we must know both sides, the good and the bad, and find the balance or trade-off}.

\begin{figure*}[htbp!]
\begin{center}
\includegraphics[width=0.75\textwidth]{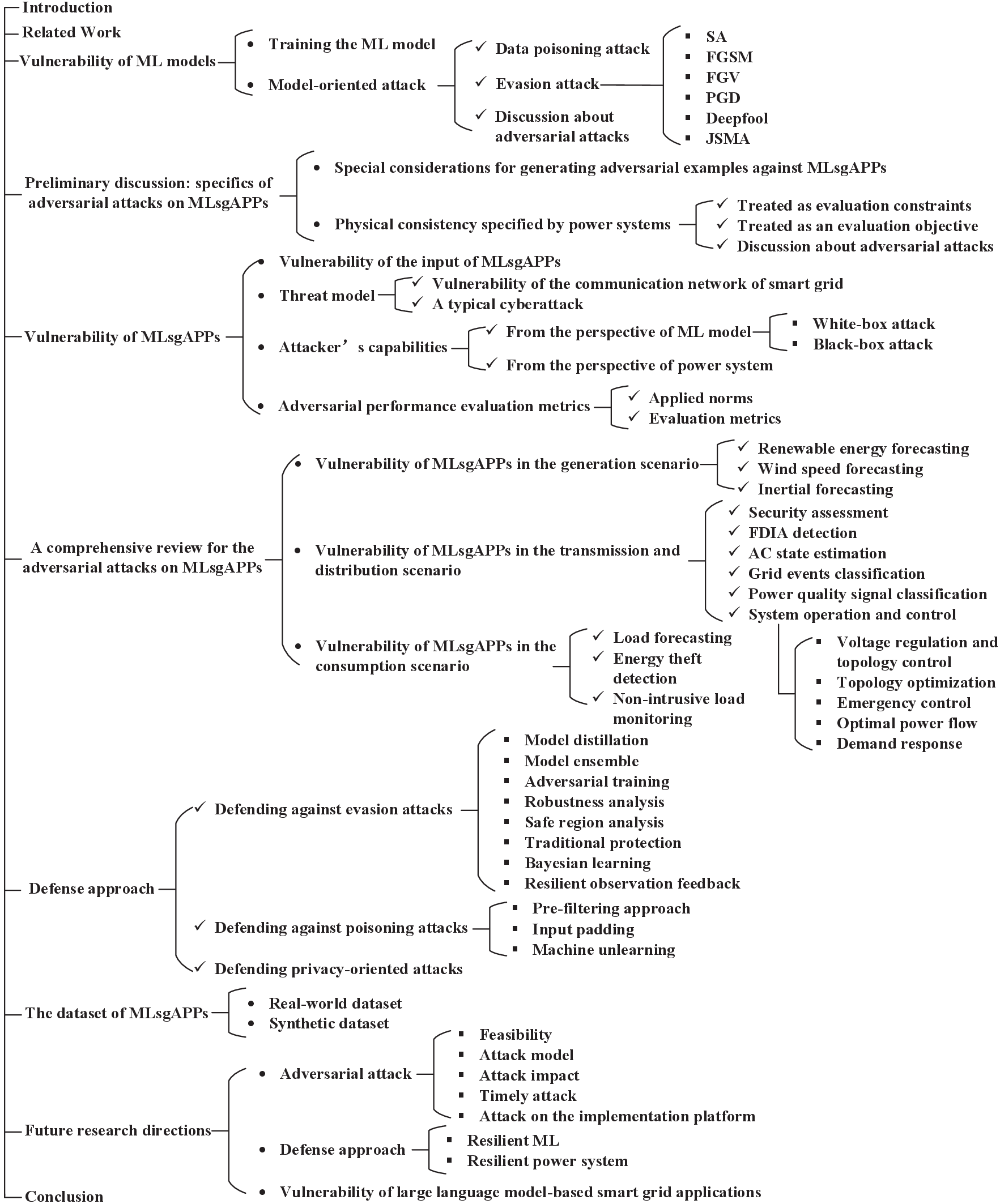}
\caption{Outline of the review on the vulnerability of MLsgAPPs.}\label{fig:outlinereviewpaper}
\end{center}
\vspace{-0.5cm}
\end{figure*}

%The intrinsic difference is that the power data are unstructured and contain complicated physical information related to the machine and electricity laws.
However, the adversarial consideration of MLsgAPP is different from that of ML approaches used in the CV domain. In power systems, the human eyes are not the only measure to observe abnormal change. The adversarial example against MLsgAPP must be perceptible to both the system operator and the deployed detector. Therefore, the approaches for constructing adversarial examples against CV applications should be modified to generate adversarial examples against MLsgAPP.  A brief review of the vulnerability of DL-based power system applications is conducted in \cite{hao2022adlsg} from the aspects of evasion and poisoning attacks and potential countermeasures. However, the number of collected papers is very limited, and thus the differences between artifacts are not pointed out. %Moreover, there lacks a discussion about the limitations, challenges, and future research directions.
Therefore, to fill the gap in highlighting the pioneer studies, we conduct a comprehensive survey of the papers on the vulnerability assessment of MLsgAPP in power grid scenarios of generation, transmission, distribution, and consumption, providing insights into this subject. In summary, our contributions are as follows:
\begin{itemize}
  \item We highlight the vulnerabilities potentially being exploited in the communication network and ML model for the real-world implementation of MLsgAPPs. The adversarial scenarios are justified using the facts from the power utilities' warnings, famous attack events, worldwide projects, and technical reports.
\item Particularly, we aim to emphasize that the adversarial attacks on MLsgAPPs are different from those against generic ML models. There are specifics in power systems reflected by the physics laws, system structure,  physical constraints, and legacy detectors.
  \item We systematically review the adversarial attacks on MLsgAPPs in three scenarios, i.e., generation, transmission and distribution, and consumption. We take an in-depth survey about the research on assessing the vulnerability of MLsgAPPs and discuss their limitations. We also summarize the countermeasures according to the methods used to enhance the security of the ML model and power data.
  \item More importantly, we provide future research directions for both the attack and defense regarding MLsgAPPs. For an attack, the feasibility, attacker's knowledge, impact analysis, and attack timing are key factors contributing to designing more stealthy, impactful, and practical attacks. For defense, the resiliency of the ML model and power system is critical to secure the use of MLsgAPPs. We also discuss the vulnerability of large language model-based smart grid applications.
\end{itemize}
The remainder of this paper is organized as follows. We first review the related surveys in Section \ref{section:relatedwork}. Then, we show the vulnerability of ML models in Section \ref{section:vulnerableML}. The preliminary discussion about the specifics of adversarial attacks against MLsgAPPs is given in Section \ref{section:difference}. Then, the adversarial attacks on MLsgAPPs are reviewed in detail in Section \ref{section:generations}. The countermeasures are summarized in Section \ref{section:defensiveapp}. Section \ref{section:datasetpowersystem} collects the dataset used for assessing the vulnerability of MLsgAPPs. The future research directions are provided in Section \ref{section:futuredirection}. Finally, Section \ref{section:conclusion} concludes the paper. The outline of this survey is given in Fig. \ref{fig:outlinereviewpaper}.

\section{Related Work}\label{section:relatedwork}
The susceptibility of ML has been attracting great attention due to the popular use of ML approaches in different areas.
For example, the use of ML in the cybersecurity field has seen an increasing trend due to the requirement of developing automatic detection and analyzing systems such as intrusion detection, banking fraud detection, spam filtering, and malware detection \cite{martins2020amlintru,liu2021artintru}. However, the ML-based cybersecurity enhancement system faces adversarial attacks \cite{zhou2019amlintrusion}. In this domain, the unique challenge for implementing end-to-end adversarial attacks lies in that the adversarial executable file must preserve the malicious functionality after modifications \cite{rosenberg2021amlcsd}. The ML approaches are also increasingly used for the tasks of wireless communication, cyber-physical systems, biometric recognition systems, and self-driving cars \cite{adesina2022amlwireless}. According to \cite{olowononi2021aisecuritycps}, the resiliency of ML approaches is important for them to be used to protect CPSs from cyberattacks. But it also reveals that compromising ML approaches in CPSs poses a challenge for attackers to seek ways to combat. Besides, the adversarial ML is comprehensively reviewed for image classifications in scenarios of biometric recognition systems and self-driving cars \cite{machado2021amlic}.

Recently, there are many survey papers about adversarial ML in the areas such as computer vision \cite{haochen2020autocom}, speech recognition \cite{hu2019amlspeech}, autonomous vehicles \cite{qayyum2020avaml}, cybersecurity \cite{zhou2019amlintrusion}, and natural language processing \cite{zhang2020nlpreview}. In \cite{jiang2021app}, the authors take an overall review of the adversarial issues in different systems. There are also generalized surveys collecting research papers about the attack and defense approaches of the adversarial ML \cite{qiu2019adreview,liu2018amlaccess, wiyatno2019amlsrevi}.

Overall, we compare the related review papers in Table \ref{tab:Advantagescompare}. These papers are compared from the aspects of the domain, whether reviewing the attack and defense approaches or not, whether considering the domain-specific properties or not, and whether considering the system-level vulnerabilities. With the growing interest in the use of ML for power systems, the significance of the vulnerability of MLsgAPPs is becoming more prevalent. However, none of the above-mentioned survey papers covers the topic of adversarial issues of MLsgAPPs. Our survey fills this gap by conducting a comprehensive survey via collecting and summarizing the key results about the vulnerability of MLsgAPPs.

\begin{table*}[htbp]
      \centering
      \caption{Comparison of the review papers about the adversarial ML}
      \label{tab:Advantagescompare}
      \begin{threeparttable}
      \begin{tabular}{p{200 pt}<{\centering}p{45 pt}<{\centering}p{45 pt}<{\centering}p{80 pt}<{\centering}p{70 pt}<{\centering}}
        \toprule
        \specialrule{0em}{1pt}{1pt}% 改变行间距
        Domain & Attack & Defense  & Domain-specific consideration  & System level \\
        \midrule
        \specialrule{0em}{1pt}{1pt}% 改变行间距
       CV domain \cite{wiyatno2019amlsrevi} & \ding{52}  & \ding{52}  & \ding{56}  & \ding{56} \\
       Intrusion and malware detection \cite{martins2020amlintru}  &\ding{52}  &  \ding{52} & \ding{56} & \ding{56}  \\
      Network traffic anomaly detection \cite{liu2021artintru} & \ding{52} & \ding{52} & \ding{52} & \ding{52} \\
   Cybersecurity enhancement system \cite{zhou2019amlintrusion}&\ding{56}  &  \ding{52} &  \ding{56} & \ding{56} \\
     Cyberattacks detection \cite{rosenberg2021amlcsd}& \ding{56} &   \ding{56} &  \ding{56}  &  \ding{52} \\
Wireless communication \cite{adesina2022amlwireless}& \ding{52} &   \ding{52} &  \ding{52}  &  \ding{56} \\
  Cyber-physical system \cite{olowononi2021aisecuritycps} & \ding{56}  &   \ding{52} &  \ding{56}  &  \ding{56} \\
  Image classification \cite{machado2021amlic} & \ding{56}  &   \ding{52} &  \ding{56}  &  \ding{56} \\
   Images, text, graph \cite{haochen2020autocom} & \ding{52} &   \ding{52} &  \ding{56}  &  \ding{56} \\
   Speech recognition \cite{hu2019amlspeech} & \ding{52}  &   \ding{52} &  \ding{52}  &  \ding{56} \\
   Connected and autonomous vehicles \cite{qayyum2020avaml}  & \ding{52} &   \ding{52} &  \ding{52}  &  \ding{52} \\
    Natural Language Processing \cite{zhang2020nlpreview}  &\ding{52} &   \ding{56} &  \ding{52}  &  \ding{56} \\
Computer vision, speech recognition, and natural language processing \cite{jiang2021app} &\ding{52} &   \ding{56} &  \ding{56}  &  \ding{56} \\
Computer vision, natural language processing, cyberspace security, and the physical world \cite{qiu2019adreview} & \ding{52}  &   \ding{52} &  \ding{56}  &  \ding{56} \\
General adversarial ML \cite{liu2018amlaccess} & \ding{52} &   \ding{52} &  \ding{56}  &  \ding{56} \\
        \midrule
Smart grid (Our work) & \ding{52} &   \ding{52} &  \ding{52}  &  \ding{52} \\
       \bottomrule
      \end{tabular}
\end{threeparttable}
\end{table*}

\section{Vulnerability of ML Models}\label{section:vulnerableML}
\begin{figure}[htbp!]
\begin{center}
\includegraphics[width=0.3\textwidth]{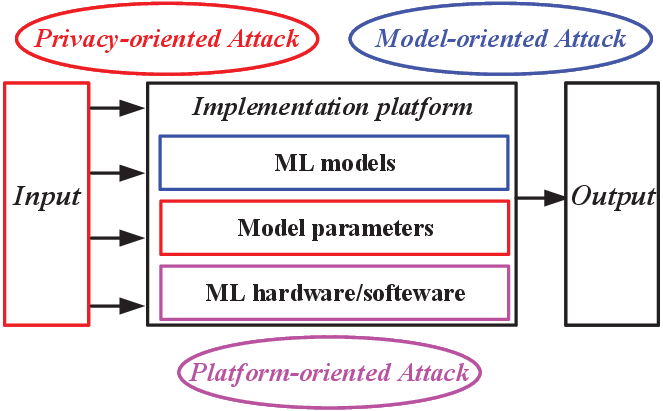}
\caption{Three types of adversarial attacks on ML. The model-oriented attack mainly focuses on the vulnerability of the ML model. The privacy-oriented attack mainly focuses on the vulnerability of the input, training dataset, and model parameters. The platform-oriented attack mainly focuses on the hardware and software on which the ML algorithms are implemented.}\label{fig:adversarilattackclassi}
\end{center}
\vspace{-0.5cm}
\end{figure}

Adversarial ML has been a well-known topic for decades. With the bloom of ML, the adversarial issues of ML raise concerns across worldwide research institutions and giant enterprises (e.g., IBM, Google, and Amazon). ML-based applications such as the PDFRATE and TEXTBUGGER have been proven easy to tamper with and manipulate \cite{rndic2014pdfrate,li2018textbugger}. The autonomous driving system faces challenges due to the weak robustness of ML models against adversarial perturbations \cite{kim2021autocyber}. Up to now, many types of adversarial attacks on ML have been discovered and introduced. Data poisoning attack contaminates the training dataset to control the output of the ML model \cite{jagielski2018poisoningatt}. The evasion attack disturbs the input with small errors to mislead the ML model's prediction \cite{biggio2013evasionatt}. The users' private data might be leaked under the model inference attack \cite{shokri2017meminfer}. The model extraction attack infers the model parameters and structures by querying the outputs given specific inputs \cite{jagielski2020extraction}. There are also other attacks including the model reverse \cite{fredrikson2015modelreverse}, backdoor attack \cite{gu2017backdoor}, and software vulnerability exploitation \cite{xiao2018exploit}.
%The authors in \cite{machado2023aml} classify the adversarial attacks into 6 aspects from different angles: attacker's influence, attacker's knowledge, security violation, attack specificity, attack computation, and attack approach, which is given in Table \ref{table:adversarialattackcalss}.
%\begin{table}[htbp!]
%\centering
%\caption{Taxonomy of Adversarial Attacks \cite{machado2023aml}}
%\label{table:adversarialattackcalss}
%%\begin{tabular}{ll}
%\begin{tabularx}{8.5cm}{p{1cm}p{3cm}XXX}
%\toprule
%\textbf{Types} &\textbf{Attack's/attacker's angle} &  \textbf{Attacks}        \\ \midrule
%Type I & Attacker's Influence &  Poisoning attack, Evasion attack         \\ \hline
%Type II & Attacker's Knowledge    & White-box attack, Black-box attack,  Grey-box attack          \\ \hline
%Type III & Security Violation   &Integrity violation, Availability violation, Privacy violation       \\ \hline
%Type IV & Attack Specificity   & Targeted attack, Untargeted attack         \\ \hline
%Type V & Attack Computation  & Sequential attack, Iterative attack        \\  \hline
%Type VI & Attack Approach & Gradient-based, Transfer/Score-based, Decision-based, Approximation-based      \\
%\bottomrule
%\end{tabularx}
%%\end{tabular}
%\end{table}
In this article, we classify the adversarial attacks into three types: model-oriented, privacy-oriented, and platform-oriented, as shown in Fig. \ref{fig:adversarilattackclassi}. The model-oriented attack has the goal of polluting or cheating the ML model, for example, the data poisoning and evasion attacks. The privacy-oriented attack aims to infer the membership in the training dataset and parameters and structure of the ML model, for example, the membership inference and model extraction attacks. The platform-oriented attack intends to exploit the vulnerabilities exposed in the hardware and software supporting the implementation of ML algorithms, for example, the software or hardware vulnerability exploitation attack. Based on the adversarial attacks on MLsgAPPs, we mainly focus on the model-oriented attack since it is significant and extensively studied. The privacy-oriented and platform-oriented attacks will be listed as future research directions.

\subsection{Training the ML model}\label{section:aimodeltraining}
Typically, the supervised ML model is trained by minimizing a parametric objective function, that is,
\begin{equation}\label{eq:modellearning}
  \min_{\bm{\theta}} ~~~h_{\bm{\theta}}(\textbf{\emph{X}}, \textbf{y}),
\end{equation}
where $h_{\bm{\theta}}$ is a loss function designed to optimize the parameter $\bm{\theta}$ (e.g., weights, activation variables, and network structure), $[\textbf{\emph{X}}, \textbf{y}]$ is compromised by paired training samples, each of which is $\emph{\textbf{x}}=\{x_i, i = 1, \cdots, n\}$ and label $y$, and $n$ is the number of selected features. After the iterative training process, the optimization problem outputs a function $f_{\bm{\theta}}(\emph{\textbf{x}})$, ideally, which maps $\emph{\textbf{x}}$ to $y$ with the model parameter $\bm{\theta}$. Normally, the output of $f_{\bm{\theta}}(\emph{\textbf{x}})$ is the prediction $\hat{y}$. The confusion matrix is constructed according to the evaluation of whether $\hat{y}$ is equal to $y$ or not. The model-oriented attack is constructed to force the learned model $f_{\bm{\theta}}(\emph{\textbf{x}})$ to output wrong predictions/classifications.

\subsection{Model-oriented attack}\label{section:modeloriented}
\subsubsection{Data poisoning attack}\label{section:datapoison}
First, we look into the vulnerability of the training process of the ML model. The collected data can be contaminated during the sampling period or the labels are flipped during the labeling process. Or the attacker just intrudes into the data server to pollute the training data. From the attacker's perspective, since the goal is to mislead the trained model to output bad predictions, the optimization direction is opposite to the learner's. That is, the process can be formulated as a bilevel optimization problem \cite{jagielski2018poisoningatt}:
\begin{equation}\label{poisoningbilevel}
\begin{aligned}
  & \underset{{D_p}}{\text{argmax}} ~~\mathcal{L}(D_t, \bm{\theta}^*) \\
  & s.t. ~~~~~~~~~~ \bm{\theta}^*\in\text{argmin}_{\bm{\theta}} ~~ h_{\bm{\theta}}(D_{\text{tr}} \cup D_p),
\end{aligned}
\end{equation}
where $D_{\text{tr}}$ is the training dataset, $D_t$ is the test or evaluation dataset, and $D_p$ is the contaminated set. The outer optimization amounts to selecting the contaminated set $D_p$ to maximize the loss function $\mathcal{L}$ defined by the attacker on the test dataset $D_t$, while the inner optimization corresponds to retraining the parameter $\bm{\theta}^*$ on a poisoned training set including $D_p$. This bilevel optimization problem is usually solved alternately. The basic idea of the data poisoning attack is to modify the distribution of the training dataset to reconstruct the mapping relationship between the input and output of the ML model.

\subsubsection{Evasion attack}\label{section:evasionatt}
Next, we introduce the evasion attack on the ML model, which is a simple but powerful attack that creates adversarial examples with imperceptible perturbations to mislead the output of ML models. Adversarial examples are crafted inputs that are close to the original values but cause the learning model to output unexpected results \cite{yuan2019adversarialexample}. The motivation to design adversarial examples can be traced back to the origin that the input and output of NN are discontinuous and inconsistent \cite{Szegedy2013rss}.

There are two kinds of formulations to compute the adversarial example. The first is to minimize the adversarial perturbation with the constraint that the prediction is wrong, that is,
\begin{equation}\label{firstformulation}
\begin{aligned}
& \emph{\textbf{r}}^* \triangleq \min_{\emph{\textbf{r}}} ~~~~ \Vert \emph{\textbf{r}} \Vert \\
& s.t.~~~f_{\bm{\theta}}(\emph{\textbf{x}}_0 +\emph{\textbf{r}} ) \neq f_{\bm{\theta}}(\emph{\textbf{x}}_0),
\end{aligned}
\end{equation}
where $\emph{\textbf{r}}$ is the adversarial perturbation added to the input $\emph{\textbf{x}}_0$ and $\Vert \cdot \Vert$ is an arbitrary norm of a vector. The second is to maximize the loss function for a target label with the magnitude of the adversarial perturbation being limited, that is,
\begin{equation}\label{secondformulation}
\begin{aligned}
& \emph{\textbf{r}}^* \triangleq \max_{\emph{\textbf{r}}} ~~~~  \mathcal{L}(f_{\bm{\theta}}(\emph{\textbf{x}}), y)\\
& s.t.~~~\Vert \emph{\textbf{r}} \Vert \leq \delta,
\end{aligned}
\end{equation}
where $\mathcal{L}(\cdot, \cdot)$ is the loss function defined by the attacker and $\delta$ is the limit of the adversarial perturbation. Note that the loss function can be defined in different forms concerning different problems.

From (\ref{firstformulation}) and (\ref{secondformulation}), the objective function is either to minimize the adversarial perturbation or to maximize the deviation of the model output. Normally, the above optimization problems are very complicated as the ML model (e.g., NN) is nonlinear and nonconvex. In \cite{Szegedy2013rss}, the adversarial attack on DNN is first introduced. The fundamental finding is that the mapping relationship between the input and output of DNN is not continuous. Thus, a small and imperceptible perturbation added to the input can mislead the prediction or classification with a high probability. To date, plenty of research has been devoted to studying the evasion attack. Most of them try to compute the adversarial perturbation accurately, efficiently, and free from being recognized, by solving problems (\ref{firstformulation}) and (\ref{secondformulation}). Algorithms have been developed to make the generation of adversarial examples more customized. We introduce some famous algorithms in the following.

\textbf{SA:} A least-effort attack is the SA that modifies the input by multiplying the original input with a scaling factor, that is,
\begin{equation}\label{eq:sascaling}
  \tilde{\emph{\textbf{x}}} = \bm{\lambda}^T\emph{\textbf{x}},
\end{equation}
where $\tilde{\emph{\textbf{x}}}$ is the adversarial example, $\emph{\textbf{x}}$ is the original input, and $\bm{\lambda}$ is a vector of coefficients.

\textbf{FGSM:} The FGSM updates the adversarial example using a one-step computation along with the sign of the gradient \cite{goodfellow2014fgsm}. It adds a constant perturbation to the original input to maximize the loss function, which is described as
\begin{equation}\label{eq:perturbations}
  \tilde{\emph{\textbf{x}}} = \emph{\textbf{x}} + \gamma \text{sign}(\nabla_\emph{\textbf{x}} \mathcal{L}(f_{\bm{\theta}}(\emph{\textbf{x}}), \bar{y})),
\end{equation}
where $\gamma$ is the perturbation limit, the operator $\text{sign}(\cdot)$ calculates the sign of a real number, $\nabla_\emph{\textbf{x}}$ computes the gradient with respect to $\emph{\textbf{x}}$, and $\bar{y}$ is the target class. It is shown that the perturbation added to each feature is the same with respect to the target label. The perturbation magnitude is controlled by $\gamma$, which cannot be too large to remain imperceptible. Since there is no constraint added to the perturbation, the adversarial input might be infeasible in the physical context, e.g., the power system scenario.

\textbf{FGV:} Different from FGSM, the FGV algorithm directly employs the original gradient value \cite{rosa2016fgv}, given by
\begin{equation}\label{eq:fgvrepresentation}
   \tilde{\emph{\textbf{x}}} = \emph{\textbf{x}} + \gamma \nabla_\emph{\textbf{x}} \mathcal{L}(f_{\bm{\theta}}(\emph{\textbf{x}}), \bar{y}).
\end{equation}

\textbf{PGD:} As a variant of FGSM, the PGD algorithm iteratively computes the adversarial example, which is given as follows \cite{madry2017tdl}:
\begin{equation}\label{eq:iterativepgd}
  \tilde{\emph{\textbf{x}}}^{k+1} = \Pi_\epsilon \{ \tilde{\emph{\textbf{x}}}^{k} + \eta \text{sign}(\nabla_{\emph{\textbf{x}}^k} \mathcal{L}(f_{\bm{\theta}}(\emph{\textbf{x}}^k), \bar{y})\},
\end{equation}
where $\Pi_\epsilon$ means projecting the adversarial perturbation within a limit $\epsilon$, $\eta$ is the step size for each iteration, and $\tilde{\emph{\textbf{x}}}^{k}$ represents the adversarial example at the $k^{th}$ iteration. The initial point is set as a random noise.

\textbf{Deepfool:} Deepfool works in an iterative manner to minimize the perturbation \cite{dezfooli2016deepfool}. It executes multiple rounds around a specific data point and outputs an adversarial example once it is misclassified. The Deepfool approach is formulated as follows:
\begin{equation}\label{eq:deepfool}
  \tilde{\emph{\textbf{x}}}^k = \tilde{\emph{\textbf{x}}}^{k-1} - \frac{f_{\bm{\theta}}(\tilde{\emph{\textbf{x}}}^{k-1})}{\Vert \nabla_{\tilde{\emph{\textbf{x}}}^{k-1}} \mathcal{L}(f_{\bm{\theta}}(\tilde{\emph{\textbf{x}}}^{k-1}), \bar{y})\Vert^2}\nabla_{\tilde{\emph{\textbf{x}}}^{k-1}} \mathcal{L}(f_{\bm{\theta}}(\tilde{\emph{\textbf{x}}}^{k-1}), \bar{y}).
\end{equation}

\textbf{JSMA:} To improve the effectiveness of the adversarial attack, the JSMA constructs a saliency map using the Jacobian matrix of the victim NN model to find out the most sensitive features \cite{papernot2016limit}. It contains 3 steps:
\begin{itemize}
  \item Compute forward derivative:
  \begin{equation}\label{eq:step1}
    \nabla_{\emph{\textbf{x}}} = \frac{\partial f(\emph{\textbf{x}})}{\partial\emph{\textbf{x}}_i} = \left [  \frac{\partial f_j(\emph{\textbf{x}})}{\partial\emph{\textbf{x}}_i}  \right ]_{i = 1, \cdots, n, j = 1, \cdots, m},
  \end{equation}
  where $\emph{\textbf{x}}_i$ is the $i^{th}$ component of the input $\emph{\textbf{x}}$, $ f_j(\cdot)$ is the $j^{th}$ layer function of NN, and $m$ is the number of NN layers.
  \item Build a saliency map:
  \begin{small}
   \begin{equation}\label{eq:step2}
    G^+(\emph{\textbf{x}}_i, \hat{y}) = \begin{cases}
      0,\text{if $\frac{\partial f(\emph{\textbf{x}})}{\partial\emph{\textbf{x}}_i} |_{y = \bar{y}} < 0$ or $\sum_{\bar{y}' \neq \bar{y} }\frac{\partial f(\emph{\textbf{x}})}{\partial\emph{\textbf{x}}_i} |_{y = \bar{y}'} >0$}  ;\\
      -\frac{\partial f(\emph{\textbf{x}})}{\partial\emph{\textbf{x}}_i}|_{y = \bar{y}} \left \vert \sum_{\bar{y}' \neq \bar{y} }\frac{\partial f(\emph{\textbf{x}})}{\partial\emph{\textbf{x}}_i} |_{y = \bar{y}'} \right \vert,  \text{otherwise}; \\
  \end{cases}
  \end{equation}
  \end{small}where $G^+(\emph{\textbf{x}}_i, \bar{y})$ evaluates the positive relationship between $\emph{\textbf{x}}_i$ and $y = \bar{y}'$ and the negative relationship between  $\emph{\textbf{x}}_i$ and $y \neq \bar{y}'$, $\bar{y}'$ is the class selected among all classes, and $\left \vert \cdot \right \vert$ denotes the $\ell_0$ norm.
  \item Repeat this process until find the features positively correlated with the targeted class $\bar{y}$. By adding small perturbations to these features, the resulting adversarial example has a large possibility of being classified into class $\bar{y}$.
\end{itemize}

In the latest decade, there are many variants of the above algorithms to make the evasion attack more powerful. For example, the BIM \cite{kurakin2016amlatscale} is an iterative variant of FGSM. The C$\&$W \cite{carlini2017robustnn} is a gradient-based attack and the adversarial perturbation is minimized to thwart the distilling defense.
Overall, the evasion attack is either a one-step attack or an iterative attack. The former has better transferability while the latter is more likely to succeed and hard to detect \cite{kurakin2016amlatscale}.

\subsubsection{Discussion about adversarial attacks}\label{section:findingsadversarialattack}
From above, it is possible that the ML model can be fooled by the adversarial attack. The selection of the attack algorithm depends on the scenario and adversarial goal. Anyway, no matter what algorithm is used, the attacker must have a good knowledge of the input of the ML model. In the CV domain, it is clear that the input is the image that is compromised by structured pixels. However, the description of the input can be confidential information that is not public in the context of power systems. Therefore, it is a critical task for attackers to find ways to compromise the input of MLsgAPP.

\section{Preliminary Discussion: Specifics of adversarial attacks on MLsgAPP}\label{section:difference}
In this section, we highlight the differences in constructing adversarial attacks in the CV domain and power system domain.

{\begin{table*}[htbp!]
\footnotesize
\centering
\caption{The differences between the adversarial example of an image and a power-related input.}
\label{table:difftwodomains}
%\begin{tabular}{ll}
\begin{tabularx}{18cm}{p{3cm}p{7cm}XXX}
\toprule
\textbf{Aspects} & \textbf{CV domain} &  \textbf{Power system domain}        \\ \midrule
Elements of the input  &  Every pixel has the same meaning and does not have physical correlations with other pixels & Different power variables have different physical meanings and are correlated with each other  \\ \hline
Time-series input & The input changes suddenly if the scenario changes & The change of the input is slow considering the mechanical inertial \\ \hline
Range of the input & The input changes in a fixed range & The input range changes a lot with respect to the operating condition and scenario \\ \hline
Evaluation metric & The $\ell_0$ norm is meaningful as the change of a single pixel is imperceptible & The $\ell_0$ norm might cause a certain power variable beyond the normal physical limits \\ \hline
Knowledge of the input& It is easy to obtain complete information of the input image & It is difficult to obtain the input to MLsgAPP as they are collected from geographically dispersed meters \\
\bottomrule
\end{tabularx}
%\end{tabular}
\end{table*}}
\subsection{Special considerations for generating adversarial examples against MLsgAPPs}\label{section:specialconsideration}
The adversarial example against MLsgAPP is different from that against the image-related application in the CV domain. A typical example of adversarial examples in the CV domain is as follows. Given an image of $2000\times1000$ pixels representing a red traffic light, by changing just 1 pixel at one corner of the image, the NN classifier could misclassify it as a green light (as shown in Fig. \ref{fig:adversarialexample}). However, the only modified pixel should be changed with a large error, that is, the modified pixel should be the green color to mislead the red light to green light \cite{wu2020traficlight}. This is not applicable in power systems because every power data has its physical constraints and limits. Different power variables are correlated with each other via physics laws. Even though only one power data is modified, the input of MLpsAPP will be regarded as an outlier due to the inconsistency.

The differences between the adversarial example of an image and a power data are (also given in Table \ref{table:difftwodomains}):
\begin{itemize}
  \item Each pixel in an image carries the same significance     (each pixel can be any value within [0, 255])  and does not have physical correlations with other pixels, whereas various power variables hold distinct physical interpretations and often exhibit correlations stemming from the laws of physics.
  \item The time-series image data might have a sudden change due to the scenario change, while the time-series power data must change slowly following the inertial property.
  \item The value of each pixel is fixed to be in the range of $[0,1]$ or $[0, 255]$, while the range of the power variable can change seriously under different operating conditions due to the periodic scheduling, seasonable power injection, etc.
  \item The evaluation metric $\ell_0$ norm for the adversarial perturbation is not suitable for the power variable since this norm might induce large errors in some power variables, which can be easily detected by legacy detectors or observed by system operators.
  \item It is easy for the attacker to obtain complete information about an image and can tamper with any pixel or frame. However, it is difficult for the attacker to know complete information about the power data. The background knowledge of images is in abundance, while the insight about power data sometimes is not available.
\end{itemize}

In power systems, the SCADA/PMU network collects the raw data from field sensors that might be disturbed by environmental noises and/or electronic faults. Therefore, the sensor measurements gathered by the control center are not reliable. The SE is responsible for reconstructing the system states from unreliable sensor measurements to provide reliable data for power system applications. The state estimation is in the loop between the data acquisitions and power system applications. Considering the adversarial attacks on MLsgAPP, the attacker compromises the communication network to inject adversarial perturbations into the sensor measurements. Due to the BDD, being equipped with SE, the adversarial examples should remain stealthy against BDD when they are transmitted to the control center \cite{liu2011false}.

Therefore, the adversarial examples against MLsgAPPs must consider the following factors:
\begin{itemize}
  \item \textbf{\emph{Stealthiness}}: The adversarial perturbation added to the sensor measurement should bypass the error-checking mechanisms such as human inspection and BDD, rather than misleading the ML model.
  \item \textbf{\emph{Timeliness}}: The construction process of the evasion attack cannot be computationally intensive. Otherwise, the attack would miss the most suitable time slot or trigger the error-checking alarm.
  \item \textbf{\emph{Physical consistency}}: The adversarial examples must follow the physics laws (e.g., Kirchhoff's Current and Voltage Law), power balance, and power limits explicitly satisfied in power systems.
  \item \textbf{\emph{Impact}}: The adversarial attack should cause a certain impact on the operating condition and/or the system's operation and control. Otherwise, the attack is ineffective although the ML model is fooled.
\end{itemize}

\begin{figure}[htbp!]
\begin{center}
\includegraphics[width=0.45\textwidth]{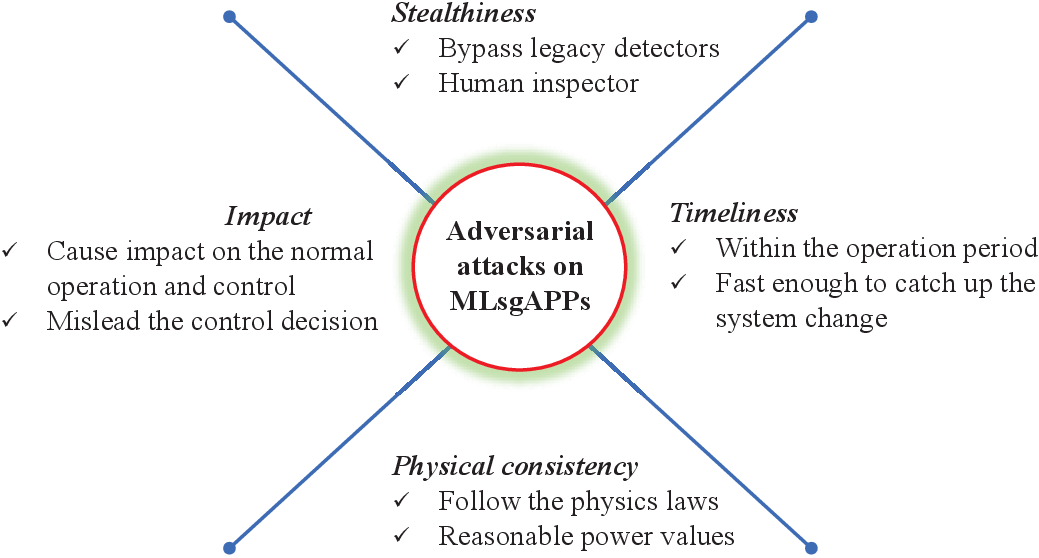}
\caption{Special considerations for constructing adversarial attacks on MLsgAPPs.}\label{fig:adversarialexamplesmlsgapp}
\end{center}
\vspace{-0.5cm}
\end{figure}

As shown in Fig. \ref{fig:adversarialexamplesmlsgapp}, these four factors are special considerations for constructing adversarial attacks on MLsgAPP. These unique requirements are deeply correlated with the characteristics of power systems.

\subsection{Physical consistency specified in power systems}\label{section:pspc}
The adversarial examples against MLsgAPPs must be constructed to meet the constraints brought in by the power system's physical consistency. The adversarial examples that violate the power limits and BDD can be easily detected because they are not consistent with the inherent properties of power systems. Therefore, it might be trivial to design adversarial attacks only from the perspective of the ML model. The adversarial examples against MLsgAPPs should not only be subject to misleading constraints but also the constraints specified in power systems. There are methods to address this issue.

\subsubsection{Treated as evaluation constraints}
First, the constraints are formulated according to the attacker's knowledge, physics laws, and execution time \cite{li2021conaml}. They are treated as linear equality and inequality constraints in the optimization problems (\ref{firstformulation}) and (\ref{secondformulation}) for generating adversarial examples. Specifically, the constrained problem is solved as follows. First, the attacker's knowledge is regarded as an unknown factor by testing if an adversarial perturbation is effective under different adversarial settings. Second, the gradient-based approach is used to generate adversarial examples. Third, the equality constraint is addressed by projecting the adversarial perturbation from the gradient value to the null space of the coefficient matrix, while the linear inequality constraint is addressed using an if-then logic to filter out unsatisfied adversarial perturbations. However, the convergence and generality of the proposed algorithm are not guaranteed. As the iterative process is applied to generate the adversarial example, the timely issue becomes serious when the system size is very large.

From a numeric perspective, Venzke \textit{et al.} \cite{venke2021verifynn} evaluates the safe region for the NN-based static and dynamic security assessment. The safe region is defined as a set of inputs that cannot be modified within a certain range to mislead the prediction/classification results. An interesting finding drawn from this work is that the safe region is affected by the power balance constraint for some tested operating points, while the power balance constraint does not affect the safe region if the constraints of generation and load limits are considered.

To be more practical, the access capability and resource limitation for generating adversarial examples should be considered. The adversarial examples are generated by solving an optimization problem with constraints introduced by the BDD, perturbation magnitude, and the number of compromised measurements \cite{liu2019annsg}. Considering the DRL-based UVLS, the adversarial examples are generated with the FGSM and JSMA algorithms to meet the load and voltage constraints \cite{wan2022drllcps}. More constraints such as the attack objective, power balance, power limits, and stealthiness are taken into account in \cite{zeng2022pcvadrl}.  However, the added constraints will largely increase the complexity of computing the adversarial example, which has time issues and can make the solution infeasible at some operating points.

On the other hand, from a systematic view, it is shown that the constraints of power balance, power limits, and BDD can make the ML-based security assessment model more robust against adversarial examples \cite{zhang2022pcrvisaps}. This is an important finding indicating that the attacker needs to pay more effort to successfully carry out the adversarial attacks. Due to the close-coupling relationship among power variables, if some sensor measurements cannot be modified, the attacker may fail to construct the adversarial examples.

\subsubsection{Treated as an evaluation objective}
Second, the specifics are considered as an objective to be minimized or maximized. For the task of ML-based false data injection detection, Liu \textit{et al.} \cite{liu2021gandiam} pointed out that there are limitations for the attacker to access the sensor measurements and bypass the BDD. Therefore, each adversarial perturbation is forced to be small to make the residue smaller than the BDD detection threshold. The normalized cross-correlation and peak signal-to-noise ratio are used as metrics to quantify the similarity between the modified and normal data. However, since only a limited number of sensor measurements are compromised, the attack vector might be detected by the BDD that uses the complete information of the system.

The BDD is also considered as an objective for generating adversarial examples against the ML-based stability assessment \cite{liu2021gandiam}. Since the attacker's capability is limited, a nonconvex, nonlinear, and mixed-integer optimization problem is formulated to minimize the adversarial perturbation and maximize the attack successful rate, about the limited number of attackable nodes, the bounds of perturbations, and the potential impact on the control consequence. However, this complicated optimization problem can only be solved using a heuristic search algorithm that is computationally intensive.

From above, it seems that the adversarial perturbation can only bypass the BDD if it is very small. However, it depends on how to define ``small" for the adversarial perturbation. Different evaluation metrics have different characteristics to quantify the size of the adversarial perturbation. To bypass the BDD, the adversarial perturbation is projected onto the nullspace of the coefficient matrix of the residue \cite{li2021adnn}. More practical, anomaly detection and error-checking mechanisms are considered for the stealthiness of the adversarial attack \cite{wang2021blackboxnilm}.

\section{Vulnerability of MLsgAPPs}\label{section:aml}
\begin{figure*}[htbp!]
\begin{center}
\includegraphics[width=1\textwidth]{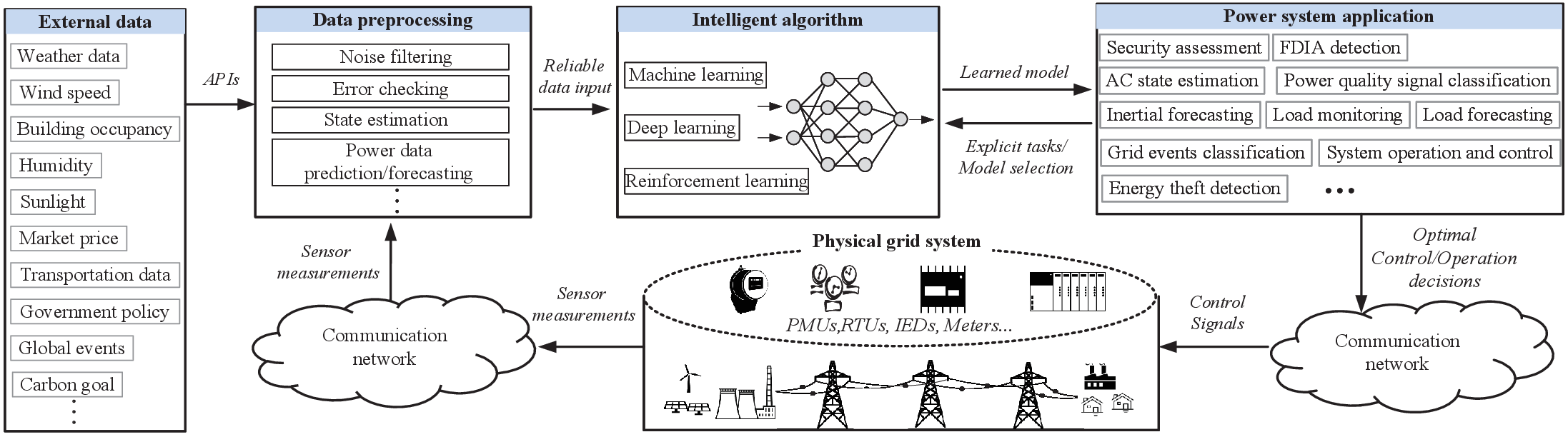}
\caption{The implementation framework of MLsgAPPs.}\label{fig:implementationstructure}
\end{center}
\vspace{-0.5cm}
\end{figure*}

Due to the worldwide energy crisis and greenhouse effect, the power system sees a significant transition to a secure, smart, and low-carbon smart grid. To empower this transition, the modern power system is built with large-scale penetration of renewable energy resources, widespread deployment of power electronic devices, and deep integration of cyber and physical spaces. However, these new facilities make the power network much more complicated and bring in a lot of uncertainties, which inspire the use of ML approaches in power systems. To name a few, there are NN-based local control policies for the distributed energy resources \cite{karagia2019mlml}, NN-based security-constrained optimal power flow \cite{martinez2011nnopf}, NN-based solutions for the $N-k$ problems \cite{donnot2017mlforpsos,donnot2018fast}. It seems that the MLsgAPPs can achieve better performance in terms of extracting the hidden features from the high-dimension and complex power data \cite{Duchesne2020reviewMLapp, Wehenkel2012learningps,Glavic2019drlps}. In Fig. \ref{fig:implementationstructure}, we show the implementation framework of MLsgAPPs in the smart grid. The MLsgAPPs are embedded as software to support the data-driven functionalities required by the operation tasks.

In the literature, most MLsgAPPs focus on evaluating the accuracy, transferability, and interpretability of the ML model. The prediction performance can achieve 99.9\% by optimizing the parameters of ML models (e.g., ensemble learning or big model). However, %as disscussed in \cite{demaontis2019transder}, if the substitute model is simpler, then the model has better transferability.
the ML approaches face a dilemma in that while impressive performance can be achieved, the hard-to-explain result confuses humans. The significant rules output from these approaches, whether inspiring or misleading, are obtained from, and only from, the training data. That is, the interpretability is poor when applying these approaches in power systems. The lack of interpretability leaves the possibility of designing adversarial examples that can mislead the operator with wrong classification or inaccurate prediction results. Although more and more studies focus on improving the interpretability of ML approaches, security is still a big issue waiting to be solved for MLsgAPPs.

\subsection{Vulnerability of the input of MLsgAPPs}\label{section:maliciousinput}
Unlike the CV domain whose input is an image composed of pixels, the input of MLsgAPPs varies a lot due to the different functionalities required to run the power system.  The input can be divided into three types: conventional meter measurements, PMU measurements, and external data from APIs or third-party services. Some studies give a default assumption that the input of MLsgAPPs can be corrupted in all cases, which deal with adversarial attacks similar to those in the CV domain. However, this is not always the case. To corrupt the input, the attacker must have the capability of exploiting the vulnerabilities exposed in sensors, communication networks, substations, and control centers \cite{ieee2011unsecure}.

\subsection{Threat model}\label{section:threatmodel}
Considering the adversarial attack on MLsgAPP, the attacker needs to figure out how to intrude into the information and communication system to obtain knowledge of the target and power data. A few decades ago, it seemed difficult to compromise an almost closed system since the grid network has strict rules for outside visitors. With the coming of Industry 4.0 or the launch of other similar projects to promote the modernization of industrial systems, the power system becomes more and more open to the outside. The borders are broken and leave a bunch of opportunities for attackers \cite{musleh2020survey2020}. It is impossible to build perfect protection to prevent all attackers anymore \cite{ghosh2019surveyscadasecurity}.

\subsubsection{Vulnerability of the communication network of smart grid} 
\begin{figure*}[htbp!]
\begin{center}
\includegraphics[width=0.88\textwidth]{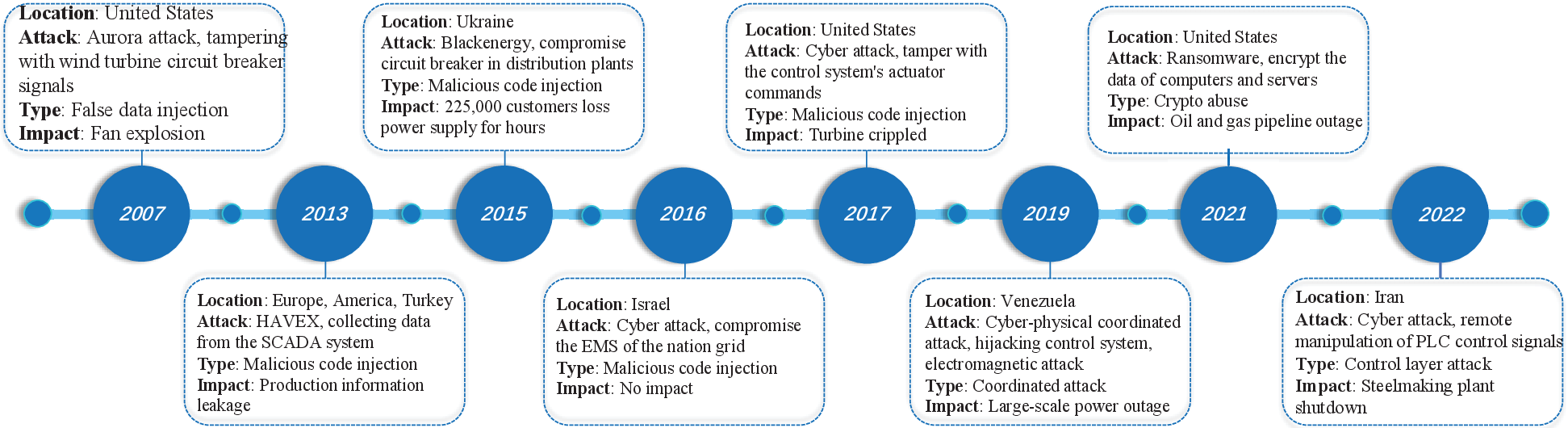}
\caption{Real-world attack events happened in power systems around the world.}\label{fig:attack_events}
\end{center}
\vspace{-0.5cm}
\end{figure*}

Pieces of evidence have shown that the attacker has the capability of compromising the power data transmitted, computed, stored, and shared in the information and communication systems. To resist illegal accesses, firewalls are commonly deployed over substations, borderlines between the production and corporation network, and interaction zones between field sites and control rooms. However, the firewalls can be overcome due to the misuse of the configuration of the bypassing whitelist \cite{zhang2019firewall1,sasaki2022rmdsecurity}. The commercial communication protocols follow a transmission routine that is well-documented and available on the producer's website. Taking advantage of the known interaction process, the execution of the man-in-the-middle attack is possible according to the explicit messaging rules from the document \cite{nawrocki2020icsin}. What's worse, the energy management system is not secure anymore. The attacker can trace back to the management system through the access points exposed in the communication network. The malware inserted into Mumbai's energy management system \cite{indian2020powergridattack} has verified this vulnerability that an attacker can breach the data collection system and steal and manipulate the production data. In 2021, the Ransomware REvil attacked the renewable energy company to steal the critical operation data.\footnote{https://www.trendmicro.com/vinfo/us/security/news/ransomware-spotlight/ransomware-spotlight-revil
} The real-world attack events that happened in recent years (as shown in Fig. \ref{fig:attack_events}) also reveal the fact that the smart grid is vulnerable to cyberattacks.

Moreover, the mentioned vulnerabilities are not only theoretical but also would happen in practice. For example, The NESCOR lists the cyber failure scenarios that might occur in the AMI and WAMPAC systems \cite{nescor2019attackinstance}. The published reports show that the smart meter can be remotely accessed by the installed malware (AMI.3), the weak encryption used in the communication network is vulnerable to the password cracking attack (AMI.24), the sensor measurements can be manipulated due to the authentication compromise in data acquisition units (WAMPAC.4), and that the local sensor measurements transmitted through the intersubstation links can be corrupted as well (WAMPAC.11). The BlackEnergy attack on Ukrainian power grid \cite{case2016ukraine} validates that the concerned cyberattacks would happen in real-world power systems. It shows that the spear phishing email with a malicious attachment can gain access to the business network of the power utility. The SCADA system is also vulnerable in that the attacker can open breakers at substations and the communication network is vulnerable in that the attacker can obtain the necessary knowledge and find entry points to take over part of operation devices \cite{yan2012smartgridcomm}. The well-designed cyberattacks support the injection of wrong control signals into substations and external data collected from third-party services.

\subsubsection{A typical cyberattack} A typical cyberattack on the smart grid is the FDIA. A variety of FDIA strategies have been proposed under different scenarios and adversarial settings. If the attacker has complete information on the system topology, line parameters, and sensor measurements, a general framework is proposed to design stealthy FDIAs \cite{liu2011false}. With the same adversarial setting, the sparse FDIA is designed to minimize the attacker's effort in executing the attack (reducing the number of sensors required to be modified) while maximizing the attack impact (inducing the largest errors into the state estimates) \cite{yang2014optimization}. If the attacker has incomplete information on the system topology and line parameters, the incomplete-information-based FDIA is designed to avoid the strong assumption for the attacker's capability of obtaining full system information. Only the information of the attack region is needed to make FDIA stealthy \cite{deng2018sparsefdi}. If the attacker has zero information on the system topology and line parameters, the blind attack is designed using the data-driven approaches to infer the unknown information and then construct FDIAs that are nearly stealthy \cite{huang2013fditopologyfree}.

\begin{figure}[htbp!]
\begin{center}
\includegraphics[width=0.4\textwidth]{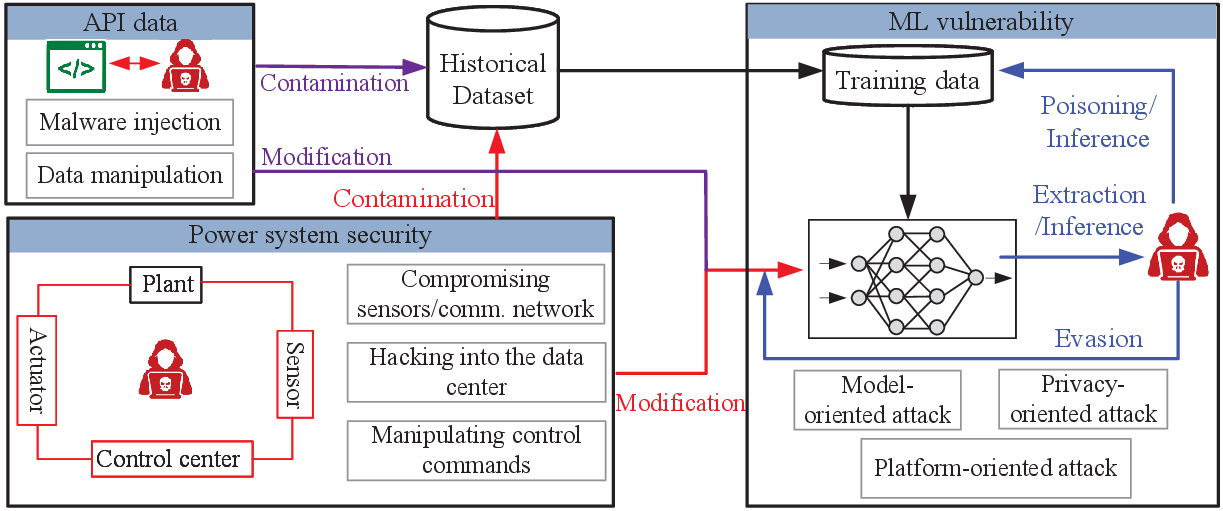}
\caption{Adversarial attacks on MLsgAPPs.}\label{fig:adversarialattackapss}
\end{center}
\vspace{-0.5cm}
\end{figure}

FDIA reveals the fact that the sensor measurements collected by the communication system can be compromised. Considering the defect of the ML model, the attacker can compromise the sensor measurements and then undermine MLsgAPPs. As shown in Fig. \ref{fig:adversarialattackapss}, we present the vulnerabilities of MLsgAPPs from the aspects of the power communication system, ML model, and external data resources. The adversarial scenarios are different due to the different adversarial assumptions and the attacker's knowledge of the ML model.

\subsection{Attacker's capabilities}\label{section:attackercapbility}
Although there are many attack models for reference, a specific attack strategy highly depends on the target scenario and attack setting. The attacker must follow what he/she has known and how many resources he/she has to determine the attacking route and method. We summarize the settings for the adversarial attacks on MLsgAPPs from the perspective of the attacker's knowledge of the ML model and the power system.

\subsubsection{From the perspective of ML model}\label{section:mlmodel}
According to the attacker's knowledge of the ML model, there are white-box and black-box attacks. Compared with the white-box attack, the black-box attack is more challenging as the attacker has less knowledge of the target. If the vulnerability of the ML model is evaluated from the defender's angle, the white-box setting is assumed to meet the thorough analysis. For example, the robustness of the ML model is evaluated to compute the lower bound of the adversarial perturbation that the ML model can tolerate.

Notice that, in most cases, the attackers' goals are the same. Regardless of the white-box or black-box attack settings, they want to mislead the output of the ML model with certain constraints. Therefore, the approaches used for constructing the white-box and black-box adversarial attacks are similar in some cases.

\begin{figure}[htbp!]
\begin{center}
\includegraphics[width=0.4\textwidth]{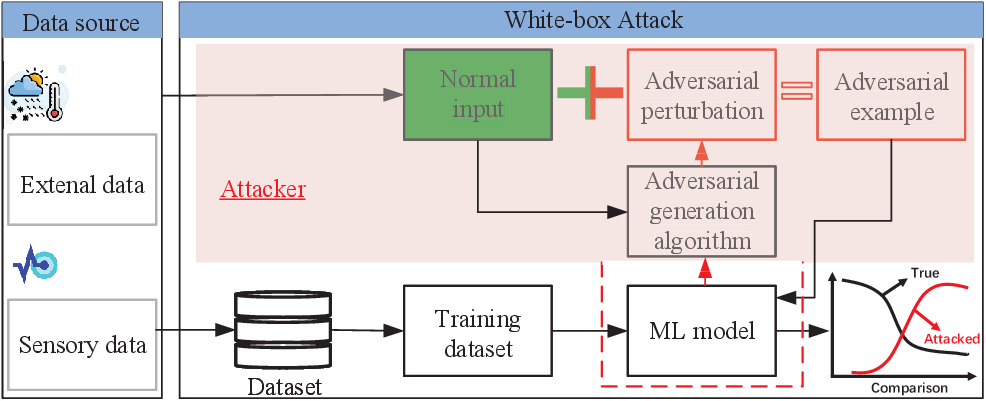}
\caption{The white-box attack against MLsgAPPs.}\label{fig:whiteboxattack}
\end{center}
\vspace{-0.5cm}
\end{figure}

\textbf{White-box attack.} Fig. \ref{fig:whiteboxattack} illustrates the white-box attack. One thing should be emphasized that, although the attacker has full information of the ML model, it does not mean that he/she can change the prediction with a 100\% successful rate. As most adversarial examples are constructed with linearized models and gradient-descent approaches, they are not perfectly accurate against the prediction process of the ML model. With the white-box assumption, the attacker is usually the malicious insider and has the capability of eavesdropping on the network traffic and hacking into the database system \cite{liu2018insiderthreats,wang2013cyber}. The white-box assumption may not always be practical, but it can help us to explore the vulnerability of MLpsAPP under the worst-case scenario, which provides guidance to constructing effective countermeasures \cite{tian2021seaann}. For the data poisoning attack, the white-box assumption provides empirical upper bounds for the performance degradation of the ML model \cite{Kammal2021ppaclsevent}. Since all parameters of the regressor are known by the attacker, the prediction error of the test data can be maximized and the adversarial perturbation can be minimized \cite{remigijus2006lipschitz}.

\begin{figure}[htbp!]
\begin{center}
\includegraphics[width=0.4\textwidth]{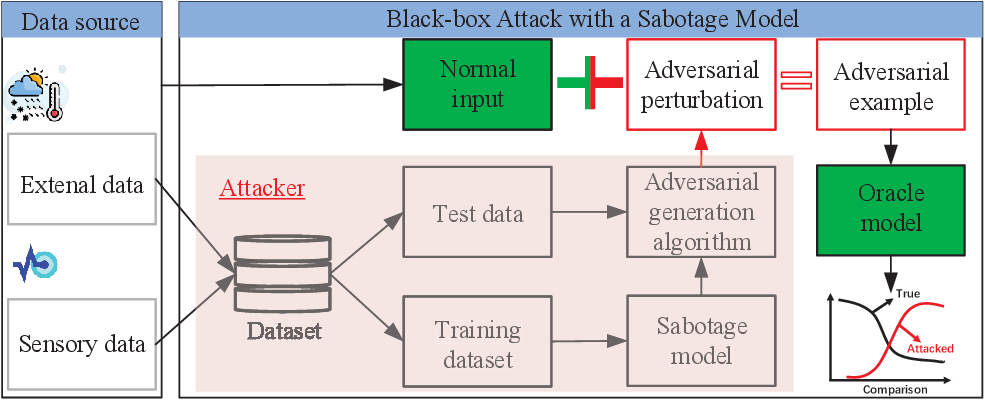}
\caption{The black-box attack by training a sabotage model.}\label{fig:adversarialattacksabotagemodel}
\end{center}
\vspace{-0.5cm}
\end{figure}

\begin{figure}[htbp!]
\begin{center}
\includegraphics[width=0.4\textwidth]{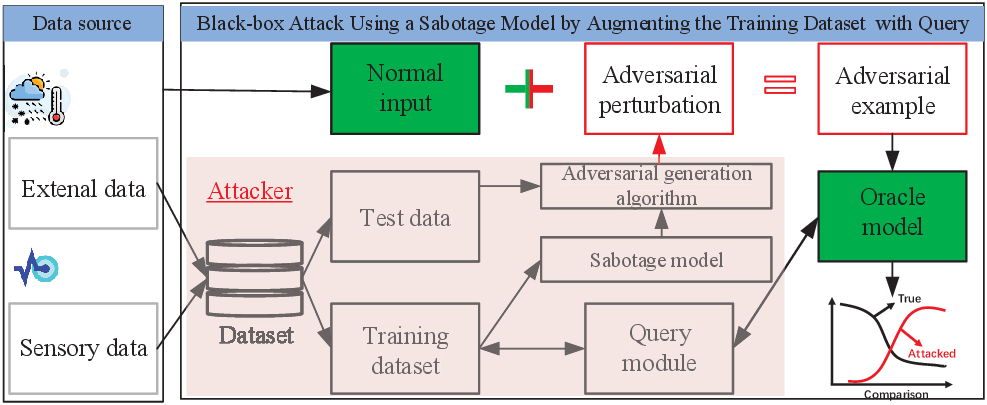}
\caption{The black-box attack by training a sabotage model with queries to the oracle model. The query is supported by some commercial applications such as the mobile APP Trickl released by the power utility company.}\label{fig:adversarialattacksqueryorcasabot}
\end{center}
\vspace{-0.5cm}
\end{figure}

\begin{figure}[htbp!]
\begin{center}
\includegraphics[width=0.4\textwidth]{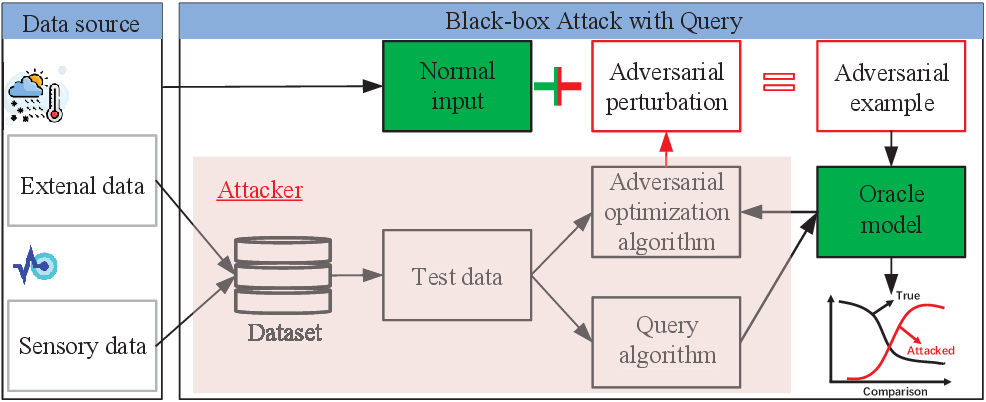}
\caption{The black-box attack by querying the oracle model.}\label{fig:adversarialattacksqueryorca}
\end{center}
\vspace{-0.5cm}
\end{figure}

\textbf{Black-box attack.}
However, the architecture of the ML model is complex and the model weights are often kept secret and thus hard to obtain by the attacker. Therefore, the black-box attack seems more practical. Under the black-box scenario, there are multiple ways to construct the adversarial attack: i) the attacker is assumed to have access to a similar or the same dataset as the Oracle model (the ML model used by the system operator) and trains a substitute model to generate the adversarial examples (as shown in Fig. \ref{fig:adversarialattacksabotagemodel}); ii) the attacker queries the oracle ML model to obtain the gradient estimation to create the training dataset and then learns a substitute model (as shown in Fig. \ref{fig:adversarialattacksqueryorcasabot}); iii) the attacker queries the oracle ML model to obtain the gradient estimation to generate the adversarial examples (as shown in Fig. \ref{fig:adversarialattacksqueryorca}). The GAN network is also located in the black-box attack setting if the Oracle ML model is not known.

As ML is increasingly provisioned as a service in big data, the power utilities might outsource the task of training the complicated model to a cloud, which sometimes is a public platform such as the Google AI platform\footnote{https://ai.google/}. Although the public cloud can protect the original data, model parameters, and structure, it can be queried remotely and is also subject to adversary attacks.

\subsubsection{From the perspective of power system}\label{section:powersysteminformation}
An interesting finding is that most existing studies about the adversarial attacks on MLsgAPPs assume a fully knowledgeable attacker who knows complete information about the power system and model input \cite{liu2019annsg,ren2022vulverifications,ren2022vulpsstvs,venke2021verifynn,zhang2022pcrvisaps,zhang2022corestability,chen2022costoriented,sayghe2020adlpsse,tian2021jae,tian2021amlfdia,sayghe2020adlpsse,li2021adnn}.
Very few studies hold the view that the attacker's capability is limited \cite{tian2021seaann,liu2021gandiam,niazazari2020attgec,zeng2022pcvadrl}.

The knowledge of the power system can be obtained by eavesdropping the communication network and compromising the operation devices. The input (e.g., sensor measurements) of MLsgAPP can be corrupted by cyberattacks such as the man-in-the-middle attack, Trojan-horse attack, and firmware modification attack \cite{wang2013cyber}. For example, the smart meter installed at the consumer's house can be easily breached and thus the power consumption data is corrupted \cite{li2020dletd,li2020searchfromfree,wang2021blackboxnilm}. The price barging process between users and power utilities can also be overheard \cite{guihai2021amlfdia}.

\subsection{Adversarial performance evaluation metrics}\label{section:evaluationmetric}
The evaluation of performance degradation is critical to assess the vulnerability of MLsgAPP. For the adversarial ML, it is important to quantify the size of the adversarial perturbation. Different metrics have been used to address this issue.

\subsubsection{Applied norms}
The well-known norms used for measuring the distance between normal data and adversarial examples are $\ell_0$, $\ell_1$, $\ell_2$, and $\ell_{\infty}$ norms. In \cite{ren2022vulverifications}, the $\ell_1$, $\ell_2$, and $\ell_{\infty}$ norms are used to analyze the bounds of the robustness of the ML-based power system stability assessment. A clear conclusion is obtained that the robustness is different using different norms. However, there is no insight about which norm has the best performance than the others. In \cite{zhou2022loadforeacting}, the adversarial perturbation added to the input is constrained by the $\ell_1$, $\ell_2$, or $\ell_\infty$ norm as the attacker's will for the ML-based load forecasting. In \cite{venke2021verifynn}, there is a provable guarantee for the input not being perturbed to the other classification using the  $\ell_{\infty}$ norm. An interesting finding is that the adversarial examples generated with the $\ell_\infty$ norm have better transferability than those generated with the $\ell_1$ and $\ell_2$ norms \cite{niazazari2020attgec}. Since the voltage magnitude and phase angle are within a very tight range, the $\ell_\infty$ norm is used to minimize the adversarial perturbation to make the adversarial attack on the NN-based AC state estimation reasonable \cite{liu2019annsg}. In \cite{ren2022vulpsstvs, zhang2022pcrvisaps, zhang2022corestability}, the $\ell_2$ norm is used to evaluate the robustness of MLpsAPP. The $\ell_2$ norm is also used to maximize the regression error \cite{remigijus2006lipschitz}. For the RL-based control \cite{chen2021safedrl}, the $\ell_2$ norm is used to limit the adversarial perturbation within a certain threshold when adding it to the state observations. Although the norms are widely used, it might not be suitable to use certain norms to evaluate the vulnerability of some specific MLpsAPPs. For example, it is said that the $\ell_0$ and $\ell_\infty$ norms are not suitable to generate adversarial examples against the CNN-based power quality classification \cite{tian2022anatt}.

\subsubsection{Evaluation metrics}
On the other hand, the impact of the adversarial attack on MLpsAPP is quantified by certain ML-related evaluation metrics. The accuracy is one of the widely used metrics to show the performance degradation under the adversarial attacks \cite{tian2022aacnnpqr,liu2021gandiam,li2021conaml,sayghe2020adlpsse,Kammal2021ppaclsevent,niazazari2020attgec,wang2021blackboxnilm,sayghe2020adlpsse,chen2022fistml}.
Considering the data imbalance, for example, if the malicious data is less than the normal data, then the confusion matrix is used to fairly quantify the impact of adversarial attacks \cite{sayghe2020adlpsse,tian2021seaann}. In \cite{li2020dletd, li2020searchfromfree, li2021adnn, guihai2021amlfdia}, the detection recall is used alone to represent the percentage of detected adversarial examples considering the stealthy requirement. Some new metrics such as the adversarial accuracy defined in \cite{venke2021verifynn} are used to compute the ratio of the correctly classified data points, for which no input adversarial perturbation can change the classification within a certain threshold.

Formally, given a MLsgAPP model $f(\cdot)$ and input $\textbf{\emph{x}}$, the minimal adversarial perturbation $\textbf{\emph{r}}$ is computed by
\begin{align}
& \delta(\emph{\textbf{x}}; f)  \triangleq \underset{\emph{\textbf{r}}}{\argmin} ~~\left\|\emph{\textbf{r}}\right\|_2 \label{op:generalformulation}\\
& \text{s.t.} ~~ \Phi (\emph{\textbf{x}}, \emph{\textbf{x}} +\emph{\textbf{r}} ) \leq 0 , \nonumber \\
& ~~~~ ~ \Psi (\emph{\textbf{x}}, \emph{\textbf{x}} +\emph{\textbf{r}} )= 0 , \nonumber
\end{align}
where $\delta(\emph{\textbf{x}}; f)$ is the robustness of $f$ at the data point $\emph{\textbf{x}}$, and $\Phi (\cdot)$ and $\Psi (\cdot)$ are functions of the equality and inequality constraints, respectively. Hence, the robustness of $f$ is calculated by
\begin{equation}\label{eq:physicsconstrained}
  \gamma_{\text{adv}}(f) = \mathbb{E}_\emph{\textbf{x}} \frac{\delta(\emph{\textbf{x}}; f)}{\left\|\emph{\textbf{x}}\right\|_2},
\end{equation}
where $\mathbb{E}_\emph{\textbf{x}}$ computes the expectation over the distribution of input and $\left\| \cdot \right\|_2$ is the 2-norm. Similarly, in \cite{ren2022vulpsstvs}, there are two robustness metrics, i.e., RII and RIC, being respectively computed for the single instance and overall model. The RII and RIC are computed by
\begin{equation}\label{robustnessmetrics}
\begin{aligned}
& \text{RII} = \Vert\delta(\emph{\textbf{x}}; f) \Vert_2, \\
& \text{RIC} = \frac{1}{N} \sum_{\emph{\textbf{x}} \in \mathcal{X}} \frac{\Vert \delta(\emph{\textbf{x}}; f) \Vert_2}{\Vert \emph{\textbf{x}}\Vert_2}, \\
\end{aligned}
\end{equation}
where $ \mathcal{X}$ is the dataset for testing. In \cite{ren2022vulverifications}, there are two bounds defined for the adversarial perturbation, which are:
\begin{itemize}
  \item Upper bound $\overline{\delta}_\emph{\textbf{x}}$: Given a specific instance $\emph{\textbf{x}}$, the adversarial perturbation $\delta_\emph{\textbf{x}}$ is always smaller than the upper bound $\overline{\delta}_\emph{\textbf{x}}$;
  \item Lower bound $\underline{\delta}_\emph{\textbf{x}}$: Given a specific instance $\emph{\textbf{x}}$, the adversarial perturbation $\delta_\emph{\textbf{x}}$ is always larger than the lower bound $\underline{\delta}_\emph{\textbf{x}}$;
\end{itemize}

Considering the specifics of power systems, a new metric is proposed in \cite{zhang2022pcrvisaps, zhang2022corestability} to compute the robustness of MLsgAPPs. If there is no feasible solution for the robustness evaluation problem, then it means that the adversarial perturbation violates the constraints specified in power systems. The corresponding data point is defined as the robust data point, which cannot be perturbed to meet all constraints (e.g., misleading, power balance, power limit, and BDD). The number of robust data points is represented by the RoC metric, that is
\begin{equation}\label{eq:tvaluecalculation}
   \text{RoC} = \frac{\text{The number of robust data points}}{\text{The number of test data points}}.
\end{equation}

For the regression task, the MAPE is used to evaluate the performance of the regression method \cite{luo2018lfbechrobust, remigijus2006lipschitz, chen2019lfdja, liu2019annsg, chen2022fistml}, which is
\begin{equation}\label{eq:mape}
  \text{MAPE} = \frac{1}{n}\sum_{i=1}^{n}( \vert\frac{ \tilde{\emph{\textbf{x}}}_i - \emph{\textbf{x}}_i}{\emph{\textbf{x}}_i}\vert ),
\end{equation}
where $\vert \cdot \vert$ denotes the absolute value. Considering the regression task for predicting the system's inertia \cite{chen2022costoriented}, the MAPE is used but the authors also introduce the Rate metric to quantify the effectiveness of the attack and check whether the modified direction is correct or not, that is,
\begin{equation}\label{eq:rate}
  \text{Rate} = \frac{\text{MAPE}_{\text{attack}} - \text{MAPE}_{\text{normal}}}{\text{MAPE}_{\text{normal}}},
\end{equation}
where $\text{MAPE}_{\text{attack}}$ and $\text{MAPE}_{\text{normal}}$ are computed with and without the adversarial attack. A new loss function $\mathcal{F}$ is defined in \cite{zhou2022loadforeacting} to measure the deviation of the prediction of the learned regressor.

With the RL algorithms, the reward is usually used as a metric to measure the performance of the output control action \cite{chen2021safedrl}. There are four metrics designed to evaluate the attack effect of the DRL-based emergency control, which are effectiveness, stealthiness, instantaneity, and transferability \cite{wan2022drllcps}. In \cite{zeng2022pcvadrl}, the attack impact is evaluated using the EPD and EPDR \cite{zheng2021vldrlfirst}, given by
\begin{equation}\label{eq:rlmetric}
\begin{aligned}
  & \text{EPD} = \frac{1}{M} \sum_{i=1}^M \pi_i(s'_i)[R_i(s_i\mid \theta^v) - R'_i(s'_i\mid \theta^v)] \\
  & \text{EPDR} =\frac{1}{M} \sum_{i=1}^M \pi_i(s'_i)[1 -R'_i(s'_i\mid \theta^v)/R_i(s_i\mid \theta^v)]  ,
\end{aligned}
\end{equation}
where $pi_i(s'_i)$ is the probability of $i^{th}$ abnormal state $s'_i$ to happen, $R_i(s_i\mid \theta^v)$ and $R'_i(s'_i\mid \theta^v)$ are the control rewards before and after the attack, respectively, and $M$ is the number of states. Since the impact of the adversarial attack is specified in the power system, the power grid-related metrics are used, for example, the ramping variation of net demand, the 1-load factor, average daily peak, net electricity consumption, and carbon emissions \cite{zeng2022resiliendrl}.

\subsubsection{Takeaways}\label{section:stata}
The norms are usually used in the constraints or objective function to measure the distance between the original and expected values. For the other metrics, they are used to evaluate the impact of the adversarial attacks on the performance of MLsgAPPs. The norms can be used for both the classification and regression tasks, while the performance evaluation should use different metrics. For the classification task, it is suitable to use the metrics such as accuracy, recall, precision, and F1-score. For the regression task, it is suitable to use the metrics such as the MAPE, EPD, and EPDR. To analyze the robustness of MLsgAPPs, it is better to use the metrics such as $\delta(\emph{\textbf{x}}; f)$, $\gamma_{\text{adv}}(f)$, RII, RIC, $\overline{\delta}_\emph{\textbf{x}}$, and $\underline{\delta}_\emph{\textbf{x}}$.

\section{A comprehensive review for the adversarial attack on MLsgAPP}\label{section:generations}
\begin{figure*}[htbp!]
\begin{center}
\includegraphics[width=0.7\textwidth]{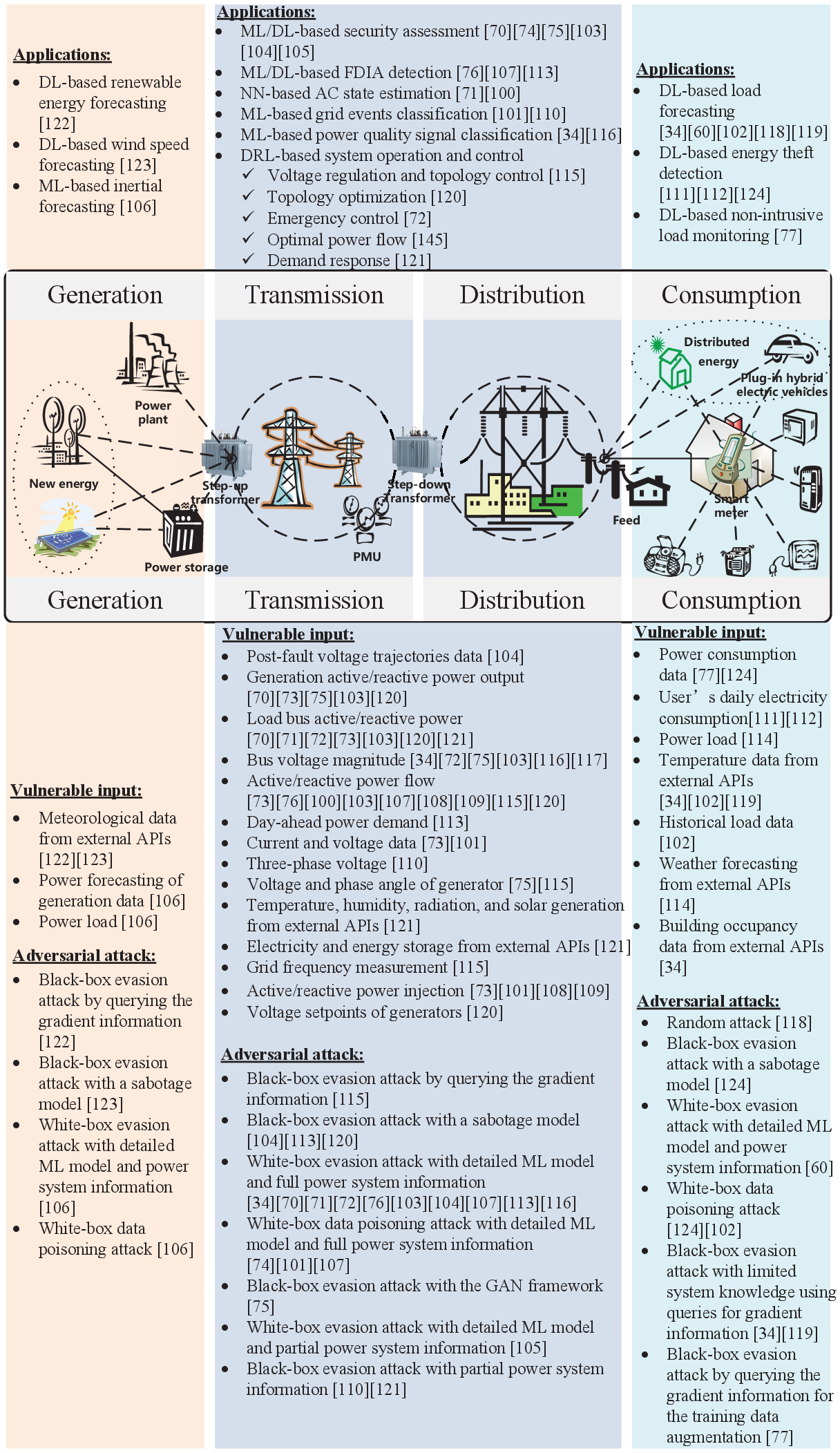}
\caption{An overview of the adversarial attacks on MLsgAPPs. In generation, studies are mainly about the vulnerability of ML-based power data forecasting. In transmission and distribution, ML is used for the optimization, operation, and control applications. In consumption, ML is used for improving the performance of load-related applications. Both the evasion and poisoning attacks can be launched to adversely affect the performance of MLsgAPPs in these scenarios.}\label{fig:adversarialattackoverviewattimpact}
\end{center}
\vspace{-0.5cm}
\end{figure*}

In this section, we review the papers about the vulnerability of MLsgAPPs in the scenarios of generation, transmission, distribution, and consumption. Although adversarial ML has been studied for around a decade, the adversarial issues of MLsgAPPs were caught in sight just several years ago. An overview of the reviewed papers is shown in Fig. \ref{fig:adversarialattackoverviewattimpact}. In each scenario, we summarize the target applications, vulnerability input, and approaches for constructing the adversarial attacks. A more detailed description of the papers is given in Table \ref{table:modelinputvulne}, where the attack impact is presented according to the experimental or simulation result obtained in each paper.

{\begin{table*}[htbp!]
\footnotesize
\centering
\caption{The detailed description of the adversarial attack on MLpsAPP in the existing paper.}
\scalebox{1}{\label{table:modelinputvulne}
%\begin{tabular}{ll}
\begin{tabularx}{18cm}{p{1.5cm}p{4cm}p{4cm}XX XX}
\toprule
\textbf{Stage} &  \textbf{Target Application}    &  \textbf{Modified Input}    &  \textbf{Attack Consequence}  \\ \midrule
 & DL-based renewable energy forecasting \cite{ruan2023vulnerability} &   Meteorological data & Under the attack by modifying the meteorological data with a 10\% error, the MAE of the PV power forecast is more than 3 times larger than that without the attack and the average economic loss is more than 700\$/10-min. \\ \cline{2-4}
  \multirow{3}{1.5cm}{\textbf{Generation}} & DL-based wind speed forecasting \cite{yang2023adversarial} &   Meteorological data &  By modifying the meteorological data with a 5\% error, the wind speed prediction deviates from the true value more than 60\% for the 1-hour, 2-hour, and 3-hour ahead forecast during any season.\\ \cline{2-4}
 & ML-based inertial forecasting \cite{chen2022costoriented} &   Power forecasting of generation data and power load & By modifying the power forecasting of generation and load with a 5\% error, the MAPE of the inertial forecasting is 2-3\% larger than the normal value.\\ \midrule
   & ML-based voltage stability assessment \cite{ren2022vulpsstvs} &   Post-fault voltage trajectories value & The assessment accuracy decreases from more than 95\% to lower than 10\% and 20\% after the white-box and black-box attacks, respectively. \\ \cline{2-4}
 & ML-based transient stability assessment \cite{ren2022vulverifications} &  Generation active/reactive power output, load bus active/reactive power, bus voltage magnitude, active/reactive power flow &    The assessment accuracy drops more than 20\% if the adversarial perturbation is beyond the lower bound with $\ell_\infty$ norm. \\ \cline{2-4}
 & DL-based N-1 security/small signal stability assessment \cite{venke2021verifynn} &   Power generation and load& The adversarial perturbation increases 3.5\% leading to 1.3\% misclassification of test operating points.  \\ \cline{2-4}
 & DL-based ac state estimation \cite{tian2021seaann} &  Active/reactive power flow & The root-mean-square error (RMSE) increases more than 10 times after the scaling attack based on the sensitivity analysis. \\ \cline{2-4}
& NN-based ac state estimation \cite{liu2019annsg} & Active/reactive power flow &  The attacks bypassing rate is more than 80\% and the relative error of the estimated result reaches more than 10\%.  \\ \cline{2-4}
 & DL-based FDIA detection \cite{sayghe2020adlpsse} &  Power flow & The FDIA detection rate drops from 99\% to lower than 50\%. \\ \cline{2-4}
 & DL-based FDIA detection \cite{tian2021jae,tian2021amlfdia} & Active power flow and power injection & The attack detection performance degrades 60\% after the attack  \\ \cline{2-4}
& DL-based FDIA detection \cite{li2021adnn} & Power flow &  The attack detection accuracy decreases from 98\% to 50\%. \\ \cline{2-4}
\multirow{21}{1.5cm}{\textbf{Transmission \& distribution}} &  ML-based FDIA detection \cite{sayghe2020adlpsse} & Active power flow and power injection  & The FDIA detection rate decreases from 99\% to 18\% in the worst-case.  \\ \cline{2-4}
& ML-based FDIA detection \cite{li2021conaml} & Power flow &  The FDIA detection accuracy drops more than 80\% after the proposed adversarial attack.\\ \cline{2-4}
 & DL-based FDI detection in demand response \cite{guihai2021amlfdia} &  Day-ahead power demand  & The FDI detection rate reduces nearly 50\% under the black-box attack.  \\ \cline{2-4}
& ML-based grid event classification  \cite{Kammal2021ppaclsevent} &  Current, voltage, buses' active/reactive power&  The classification accuracy of grid events  decreases more than 20\% after the data poisoning attack. \\ \cline{2-4}
 & DL-based grid event cause classification \cite{niazazari2020attgec} & Three-phase voltage &  The successful rate of adversarial attacks decreases from 99\% to 76\%. \\ \cline{2-4}
  &  ML-based power quality classification \cite{chen2022fistml} & Voltage & The classification accuracy decreases from 97\% to 67\% with the injection of small adversarial perturbations.  \\ \cline{2-4}
  & DL-based power quality signal classification \cite{tian2022anatt,tian2022aacnnpqr}&  Voltage & The misclassification rate of power signals can achieve 98\% and 74\% under the signal-specific and signal-agnostic adversarial attacks, respectively. \\ \cline{2-4}
  & ML-based CCT prediction \cite{liu2021gandiam}&  Generator's voltage, power angle, and active/reactive power &  The CCT increases nearly 50ms after the adversarial attack. \\\cline{2-4}
 & DRL-based emergency control \cite{wan2022drllcps}& Load and voltage  & The performance of the DRL model drops more than 30\% under the white-box recovery-target attack.     \\ \cline{2-4}
   &DRL-based AC OPF \cite{zeng2022pcvadrl}& Active/reactive generation, active/reactive load, active/reactive power injection, active/reactive power flows, and line currents  &  The performance of the DRL model decays more than 10\% under the proposed FDIAI attack.  \\ \cline{2-4}
 &Multi-agent DRL-based demand response \cite{zeng2022resiliendrl}& Date, temperature, humidity, load, electricity and energy storage, radiation, and solar generation  &  The adversarial attack can cause a 41.43\% higher ramping value of the demand curve.\\ \cline{2-4}
  &DRL-based Topology optimization \cite{zheng2021vldrlfirst}& Active power outputs, voltage setpoints of generators, loads, line status, line flows, thermal limits, and timestamps  & The performance of the DRL model under the critical adversarial perturbation drops 32.1\% and the resulting power network might be isolated. \\ \cline{2-4}
  & DRL-based voltage regulation and topology control \cite{chen2021safedrl}& Power flow, voltage, and frequency &  The adversarial attack causes line overloading for more than 1/3 of the daily test samples. \\\midrule
  & ML-based load forecasting \cite{remigijus2006lipschitz} & Temperature and historical load & The MAPE value increases from 5.22\% to more than 70\% \\ \cline{2-4}
    & DL-based load forecasting \cite{zhou2022loadforeacting}& Load and weather forecasting data & The adversarial attack make the MAPE reduces 2-6\% after adversarial attacks.  \\ \cline{2-4}
 & DL-based load forecasting \cite{chen2019lfdja}& Temperature value & The forecasting error increases from 2\% to more than 10\%. \\ \cline{2-4}
  \multirow{3}{1.5cm}{\textbf{Consumption}} &  DL-based building load forecasting \cite{chen2022fistml}&  Temperature value and building occupancy & The model's MAPE increases from 5.29\% to 25.9\%. \\ \cline{2-4}
 & ML-based non-intrusive load monitoring \cite{wang2021blackboxnilm}& Power consumption aggregated from each consumer entity &  The percentage of wrongly classified active appliances can reach more than 40\% under both white-box and white-box attacks. \\  \cline{2-4}
 & ML-based energy theft detection \cite{takiddin2021retd} & Power consumption data  & 10\% of energy thefts are undetected on average. \\ \cline{2-4}
 & DL-based energy theft detection \cite{li2020dletd,li2020searchfromfree}& User's daily electricity consumption & The detection accuracy of abnormal consumption data drops more than 50\%. \\
\bottomrule
\end{tabularx}}
%\end{tabular}
\end{table*}}

\subsection{Vulnerability of MLsgAPP in the generation scenario}\label{sec:generation}
Normally, the power system is controlled by prediction first and control in real-time. However, the integration of renewable energies makes power generation unpredictable, resulting in increasing difficulty in the management of the grid. In this case, the ML approaches have been used to keep the power balance, system stability, and minimization of generation waste. The most studied is to forecast the energy output from renewable energy power plants such as the wind power generation \cite{hossain2021predicting,li2023wind}, solar energy generation \cite{mahmud2021machine, kim2019two}, and hydropower generation \cite{djukanovic1995neural}.

\subsubsection{Renewable energy forecasting}\label{section:refeoos}
ML approaches used in the scenario of renewable energy forecasting have the purpose of maintaining the supply-demand balance with the increasing integration of distributed energy resources. However, the vulnerability of ML brings new threats to the forecasting process. Ruan \emph{et al.} \cite{ruan2023vulnerability} propose an adversarial learning attack against DL-based renewable energy forecasting by modifying the meteorological data from external APIs. The attack impact (economic cost and carbon emission) is maximized by formulating a nonlinear optimization problem. Considering a black-box attack scenario, the query is needed to obtain the gradient information. It is claimed that the adversarial attack can be generalized to the ML-based forecasting of other renewable energies. However, the input might not just be meteorological data for the other tasks. In practice, only modifying one kind of data can be easily observed, especially the weather prediction that is usually public and known by both the attacker and defender.

\subsubsection{Wind speed forecasting}\label{section:winspeedforecast}
Wind power has developed rapidly in recent years and accounts for a large part of power generation. However, the uncertainty of the weather makes it hard to accurately predict the output power of wind farms. Since the output power is related to the wind speed, the prediction of wind speed is quite important. The approaches based on NN have been widely used for extracting the wind speed change from the meteorological data. This data-driven technique can largely improve the prediction accuracy of the wind speed and the corresponding output power.

However, since wind speed prediction depends on the data from the external weather forecast system, it can be compromised by attackers exploiting the vulnerabilities of the information and communication devices. Therefore, Yang \emph{et al.} \cite{yang2023adversarial} propose an adversarial FDIA against the NN-based short-term wind speed forecasting. By corrupting the meteorological data with 1\%, the wind speed prediction can deviate more than 10\% from the true value in the black-box case. To solve the complicated optimization problem with nonlinear and non-convex constraints, a particle swarm optimization-based algorithm is developed. Although the numerical result seems convincing, the insight into the relationship between the adversarial perturbation and prediction accuracy is not provided. The slight perturbation might not be that significant in affecting the predicted result if the perturbed time and variable are not appropriately selected.

\subsubsection{Inertial forecasting}\label{section:inertialforecasting}
Inertial forecasting is an increasingly important task for guaranteeing the power system's stability due to the wide deployment of converter and inverter-enabled microgrids. It determines the frequency response after disturbances and demand-supply imbalances. Traditional inertial forecasting approaches focus on the modeling of synchronous generators. However, the power system is evolving from the isolated control room to the widely connected control units. The inertial forecasting faces much more uncertainties and needs more data to improve the accuracy. Therefore, the ML approach seems potential to be used for inertia forecasting, especially NN.

In practice, inertial forecasting also relies on the outside data (e.g., weather forecasting) collected through the third-party API. In that case, a natural thinking is that the input data of the ML model can be tampered with. Chen \textit{et al.} \cite{chen2022costoriented} propose a framework to evaluate the vulnerability of the ML-based inertia forecasting with data poisoning and evasion attacks, modifying the power data and API data. Different from the existing studies, the authors design a cost-oriented instead of an output-oriented vulnerability assessment framework. Usually, maximizing the output of the ML model might trigger the data-consistency detector. Therefore, in \cite{chen2022costoriented}, the forecasted error is minimized while the attack impact on the operation cost is maximized.  For the data poisoning attack, the poisoning strategy is formulated as a bilevel optimization problem and is solved with an alternating process using the gradient descent approach. Rather than modifying the predictions or labels \cite{jagielski2018poisoningatt, yang2017poisonatta}, the data poisoning attack is designed by perturbing partial input and making the prediction clean in the training dataset.

\subsection{Vulnerability of MLsgAPPs in the transmission and distribution scenarios}\label{sec:transmission}
Because the applications in the transmission and distribution scenarios are very similar and sometimes the same. We review the related papers as a whole.

\subsubsection{Security assessment}\label{section:dsa}
For the power system security assessment, the high computational burden to solve a sufficiently large set of algebraic equations makes the time-domain simulations hard to compile with the integration of dynamically distributed renewable energies. With the urgent need for online/real-time operations, the ML approaches have the advantage of accelerating computations while maintaining the original functionality of numerical or time-domain solutions. As the investigation for decades, the applied ML approaches for the power system security assessment include shallow learning methods such as ANN, SVM, and DT and deep learning methods such as DBN, CNN, RNN, and GAN. However, once the ML-based security assessment is corrupted, the unstable status can be wrongly predicted to be stable, hence, the preventive control action is not conducted, in consequence, the power system might suffer from cascading failure or even widespread blackout; or the stable status is falsified to be unstable and therefore the preventive control is executed to draw the system back to “normal”, causing unnecessary costs and customer interruption.

The N-1 static security issue evaluates the system's safety with the DC/AC optimal power flow problem subject to one failure (e.g., line fault, generator loss, and equipment shutdown). NN has been used to speed up the calculation and increase the number of evaluated faults. Considering the adversarial issue of this NN-based application, Venzke and Chatzivasileiadis \cite{venke2021verifynn} provide a provable guarantee for the input not being perturbed to be misclassified as an unsafe operating point within a certain threshold. The bound of the robustness is solved by formulating a MILP problem by transferring the ReLU unit into a mixed integer programming problem. The interval arithmetic and weight matrix sparsity methods are adopted to improve the efficiency of solving the MILP problem. This work assumes that all system states are possible to be modified. However, there are many limitations for the attacker to access the communication network and compromise the digital and decimal signals.

Transient security assessment considers the ability of the power system to prevent the loss of synchronism subject to sudden changes in generation, load, transmission, etc. For the NN-based transient stability assessment, Ren and Xu \cite{ren2022vulverifications} verify the NN model's robustness by modifying the theoretical results provided in \cite{remigijus2006lipschitz} to make the derived bounds satisfied in both the differentiable and non-differentiable scenarios. With the Lipschitz assumption, they relax the intractable gradient computations to calculate the local Lipschitz constant using the extreme value theory \cite{zheng2018extenva} and backward pass differentiable approximation. Although concrete analytical results are derived, the robustness bounds are difficult to compute and the Lipschitz assumption makes them not general.

The prediction of CCT of removing the fault is also critical for the system's transient stability, i.e., the rotor angle stability. Since the ML approaches are widely involved, their vulnerabilities arise from adversarial examples are analyzed in \cite{liu2021gandiam}, where a GAN framework is constructed to generate high-success-rate adversarial examples. The adversarial perturbation is minimized with the constraint that the deviation of the control output is beyond a certain threshold. However, the training process is very complicated, which might affect the efficiency of updating the attack strategy.

STVS evaluates the voltage dynamics by verifying if the voltage is within an acceptable level (i.e., over 90\% of the voltage level in the pre-fault period) or if the voltage collapses after interference. The ML approaches such as LSTM, FCNN, and BPNN have been used for STVS. The vulnerability of the DL-based STVS is revealed in \cite{ren2022vulpsstvs}. Like the traditional image-related adversarial analysis, the adversarial threat is analyzed comprehensively from the aspects of the construction of adversarial examples in white- and black-box scenarios, robustness evaluation with linear and nonlinear binary classifiers, and adversarial training for attack mitigation. However, all data collected from the field sites have physical meanings and are checked before being used, which cannot be arbitrarily perturbed as an image input. The analysis does not take the characteristics of the power system into account except for the dataset generated from the real-world power system application.

Different from above, Zhang \emph{et al.} \cite{zhang2022pcrvisaps} provide a general framework to evaluate the robustness of the ML-based security assessment considering the constraints specified in power systems. The impact of constraints such as power balance, power limits, and BDD on the robustness of this ML-based application is deeply investigated. Except for the magnitude of adversarial perturbation, a novel metric is given to measure the robustness when there are operating points that are intrinsically robust, that is, these data points cannot be perturbed to be adversarial examples due to the constraints specified in power systems. Furthermore, a robustness evaluation framework including the attacker's limited capability is given in \cite{zhang2022corestability}. In practice, the attacker can't modify all sensor measurements since the sensors are distributed on a large scale and the data is collected from different substations, which is different from the adversarial setting for the attacker in the CV domain who can modify any pixel of the image.

\subsubsection{FDIA detection}\label{section:fdiad}
The noisy sensor measurements collected by the SCADA system are used by SE to estimate the state variables (e.g., the voltage magnitude and phase angle), which are further leveraged by the control center to perform a wide range of essential functions including contingency analysis, optimal power flow, load forecasting, etc. The accuracy of SE is so important that any bad/malicious influence on it might result in making out-of-order decisions. However, the FDIA is a kind of stealthy but powerful data integrity attack that could change the state estimates of SE \cite{liu2011false}. Since the FDIA can be deeply hidden in normal data, the ML approaches have been utilized to detect FDIAs exploiting their high accuracy \cite{he2017dldetectfdia}. However, the ML-based FDIA detection methods are vulnerable to adversarial examples \cite{sayghe2020adlpsse}.

Considering the ML-based FDIA detection method implemented with ELU, the ideas of L-BFGS and JSMA are used to generate targeted adversarial examples to modify the state estimates of SE \cite{sayghe2020adlpsse}. The L-BFGS-based adversarial examples can reduce the FDIA detection accuracy to be lower than 20\% while the non-target JSMA can reduce the FDIA detection accuracy to around 10\%. However, the adversarial examples are generated without considering the specifics of power systems. It is impractical to directly apply the L-BFGS or JSMA method to construct adversarial attacks on the ML-based FDIA detection.

Moreover, random adversarial attacks sometimes are not effective due to the constraints introduced by the attacker's incomplete system information, power limits, and BDD \cite{tian2021jae,tian2021amlfdia}. It is suggested that the adversarial attack and FDIA should be jointly designed such that the adversarial perturbations could cover the sensor measurements corrupted by FDIAs. Therefore, a measurement-perturbation-based adversarial FDIA is proposed to undermine the NN-based detector with limited sensors being compromised and a state-perturbation-based FDIA is proposed by directly adding adversarial perturbations to state variables to realize perfect stealthy attacks against BDD. However, there are several concerns about these studies. First, since the state variables are usually unknown, it is almost impossible to perturb the state variables to construct the adversarial attack. Second, since the adversarial attack is generated iteratively, the construction process might fail due to the system dynamics.

Since the BDD cannot be violated, the adversarial perturbations added to the corrupted sensor measurements by FDIAs must follow the power flow model \cite{li2021adnn}. With this concern, the adversarial perturbation is projected onto the null space of the coefficient matrix of the measurement residue, which is inserted into the iterative generation process of adversarial examples. However, because the projection process is computationally intensive, the attacker might fail to successfully launch the attack as the perturbation needed to be added to the real-time measurement.

On the other hand, the data poisoning attack is also a big threat to the ML-based FDIA detector \cite{sayghe2020adlpsse}. Instead of solving a bilevel optimization problem, the flipped labels are determined by maximizing the loss caused by the adversarial training dataset on the normal SVM model. A joint optimization problem, minimizing the training loss and maximizing the adversarial loss, is formulated to obtain the adversarial model. However, this work takes a very strong assumption that the attacker has full knowledge of the training dataset.

In practice, it seems more reasonable to modify the load data since the loads are distributed on the user's side. In the distribution system, the DR scheme is implemented by combining the home energy management system and advanced metering infrastructure. The electricity consumptions are collected with commercial smart meters and sensors, which have been validated to be vulnerable against cyberattacks \cite{illera2014smartmetersnn}. Zhang and Biplab \cite{guihai2021amlfdia} evaluate the vulnerability of the CNN-based approach to detect FDIAs on the DR scheme. The attacker changes its electricity demand at certain slots to mislead other users' decisions on reporting the required demands the next day. As the power consumption peak is moved, the attacker can gain monetary benefits by buying the electricity at his/her peak with the lowest price. To bypass the CNN-based detector, the adversarial examples (i.e., the FDIA attack vectors) are generated to fool the detector into classifying them normally. Since the error injected into the attacker's electricity demand is maximized, the FGV approach is modified by adding rather than subtracting the gradient, and an iterative process is introduced to make it more confident to get adversarial examples. Since the target is the load demand on the attacker's side, the attack assumption is more reasonable and easy to realize.

\subsubsection{AC state estimation}\label{section:acstaest}
The nonlinear AC state estimation, which is used to compute state estimates with a nonlinear system model for the power system operation and control, often has a computational intensive burden and fails to converge if the system size is large. Due to the integration of fast-dynamic electronic devices, real-time AC state estimation is urgently needed. However, the sensor measurements and state variables are usually nonlinearly dependent, which sets barriers to efficiently calculating the results while guaranteeing the estimation accuracy. Nevertheless, it has been proved that the multi-layer NN can approximate any function \cite{leshno1993mlpanyfunction}. Therefore, NN has been exploited to realize the real-time AC state estimation. The historical voltage measurements are used to predict the incoming voltage, thus, transforming the computations for the nonlinear equations to an NN-based regression process. It is reported that the NN-based method speeds up several orders of magnitude of the computation time \cite{mestav2019bacse}.

However, the NN-based AC state estimation is vulnerable to adversarial examples \cite{tian2021seaann}. Based on the results obtained from the forward derivative, the most vulnerable input can be screened out using the sensitive value, which is similar to the idea of JSMA \cite{wiyatno2018jsma}. The estimation error of the NN-based AC state estimation is forced to be more than 100 times the normal value under the adversarial attack. However, the increase in estimation error can also be induced because the scaled sensor measurements do not follow the physics laws, that is, the adversarial example is no longer a reasonable operating point of the power system.

Therefore, the attackers cannot modify the sensor measurements as they will since there are access limitations and error-checking mechanisms. To evaluate the maximum impact of the adversarial attack on the NN-based AC state estimation, Liu and Shu \cite{liu2019annsg} construct a constrained optimization problem to maximize the error introduced into the state estimate with the constraints of BDD-bypassing, limited corrupted measurements, and bounded modification range. As there are nonlinear and nonconvex terms in the optimization problem, intelligent algorithms are adopted to compute the optimal attack vector. However, the algorithms are usually computationally intensive, and thereby the adversarial attack might not meet the real-time requirement for the system's normal operation.

\subsubsection{Grid events classification}\label{section:grideventclassi}
Grid events classification is used to identify the location of the root cause of an event or recognize the event types such as the transformer tap-changer operation, load switching, and capacitor bank switching \cite{samuelsson2006eventclasi}, which is critical for event-based remedial actions and preventive maintenance. As the micro-PMUs are widely deployed, the dense and massive data are collected from the field site, which enables the adoption of ML approaches to achieve accurate and real-time grid events classification \cite{yadav2019rtdetection,aligholian2021clustering}.
%The authors also reveal the features that are more prone to cause misclassification if the training dataset is poisoned.

However, since the ML-based grid events classification was not originally designed in an adversarial environment, it has vulnerabilities in its use. The adversarial instances can be constructed and injected into the training dataset, affecting the grid events classification and prompting wrong actions \cite{Kammal2021ppaclsevent}. Specifically, the data poisoning attack is constructed by solving a bi-level optimization problem in which the attacker tries to maximize the hinge loss (i.e., misclassification) and the defender intends to optimize the classification boundary. However, it needs the attacker to make plenty of effort to collect the information from the dataset. The required intensive cost and full information of the system may hinder the attacker's motivation to launch such an attack.

Therefore, in \cite{niazazari2020attgec}, a framework is established to assess the vulnerability of the CNN-based event cause analysis under the black-box scenario. Since it is very difficult for the attacker to obtain information on the ML model and system parameters, the attacker is assumed to rely on a small portion of the dataset to launch the attack. By monitoring a limited amount of voltage and current measurements, the attacker classifies the data into several events using an unsupervised approach. The dataset is labeled based on the clusters and thus a substitute CNN model is trained. The adversarial perturbation is constructed using the FGSM method and minimized by tuning the scaled coefficient. However, although a black-box scenario is assumed, it still requires the attacker to know the meaning and structure of the input signal, which is difficult since the ML-based grid event classification might use different signals.

\subsubsection{Power quality signal classification}\label{section:powerqualitysigcla}
With the fine-grained data collected from advanced sensors, the power quality can be identified by extracting the hidden features in the historical power data. Generally, power quality signals can be classified into four types: clean signal, sag signal, impulse signal, and distortion signal. The authenticity of power quality data is of vital importance as it is closely related to financial decisions, accurate and effective control, as well as operation. Once the authenticity of the power quality data is undermined, the consequent wrong decision-making might lead to load shedding and unreliable control. Since the model-based approaches need complicated computations for identifying the power quality signal, the ML methods (e.g., DT, SVM, and NN) are adopted to achieve real-time and accurate detection of unqualified data.

However, it shows that the ML-based power quality classifiers suffer from the threat of adversarial examples. Chen \emph{et al.} \cite{chen2022fistml} construct the adversarial attack based on the idea of FGSM by modifying the voltage measurements. Tian \emph{et al.} \cite{tian2022anatt} also focuses on this issue and proposes an adaptive normalized attack to undermine the power quality classifier. An interesting finding in \cite{tian2022anatt} is that traditional approaches such as the Deepfool and FGSM, will introduce large errors into the sensor measurements that can be easily detected. Therefore, an adaptive normalized attack is proposed to bind the adversarial perturbation by scaling the size but maintaining the direction. Besides, as the signal-specific adversarial examples have poor transferability to the other types of power signals, a signal-agnostic approach is proposed to avoid computing the adversarial perturbation for each specific input. However, the universal perturbation is generated by testing all samples if the added perturbation causes misclassification given a predetermined probability threshold. Besides, the adversarial examples are generated in an iterative way, which will affect the efficiency of launching the adversarial attack.

\subsubsection{System operation and control}\label{section:operationcontrol}
In power systems, some tasks need hard time constraints for the system-wide operation. This is a complicated problem since it needs to achieve satisfying performance in time scale, stability, and economics. However, the traditional model-based control and optimization approaches do not have the capability of modeling the complex system dynamics precisely, losing accommodation of distributed renewable generations, and slow response to sudden events and disturbances \cite{ernst2004rlpowe}.
Recently, the DRL approaches have been widely used for electric vehicle scheduling, demand response, load frequency control, emergency control, topology optimization, etc. A review of the DRL-based power system control and operation can be found in \cite{zhang2018drps}.

As a time-series process, the power system control problem can be modeled as a decision-making process based on the Markov decision process \cite{zhang2019drlpsapp}. The system operation and control are usually formulated as \cite{chen2021safedrl}:
\begin{equation}\label{eq:systemoperationandcontrol}
\begin{aligned}
  & \min_{\emph{\textbf{a}}(t)} ~~~ \sum \mathrm{C}\left( \emph{\textbf{o}}(t), \emph{\textbf{a}}(t)\right) \\
  & s.t. ~~~~ \emph{\textbf{o}}(t + 1) = \varphi \left(\emph{\textbf{o}}(t), \emph{\textbf{a}}(t), \emph{\textbf{u}}(t) \right); \\
  & ~~~~~~~\phi\left(\emph{\textbf{o}}(t), \emph{\textbf{a}}(t), \emph{\textbf{u}}(t) \right) = 0;\\
  & ~~~~~~~\chi\left(\emph{\textbf{o}}(t), \emph{\textbf{a}}(t), \emph{\textbf{u}}(t) \right) \leq 0,
\end{aligned}
\end{equation}
where $\emph{\textbf{o}}(t)$ represents the observations (e.g., power flow, voltage, and grid frequency) at time $t$, $\emph{\textbf{a}}(t)$ denotes the control action (e.g., generation dispatch, load shedding, battery charging, and line switching) on the power system at time $t$, $\emph{\textbf{u}}(t)$ represents the external/environment input (e.g., sudden demand and intermittent renewable generation) at time $t$, $\varphi(\cdot)$ represents the function of system dynamics, $\phi(\cdot)$ denotes the power balance, $\chi(\cdot)$ represents the power limitation, and $\mathrm{C}(\cdot, \cdot)$ is the cost function for the control and operation. The goal of the system operator is to minimize the operation cost by optimally selecting the action sequences. The difficulty in solving this optimization problem is introduced by the constraints as there are different time-scale dynamics, complex structures, and highly flexible plug-in and out renewable sources and demands. The ML approaches have the capability of smoothing the complexity and uncertainty by extracting the hidden behaviors and relations in the dynamic process, especially the cutting-edge DRL, which has the great advantage of conducting real-time operation and control. It interacts with the environment to learn the dynamic optimal decisions in sequential, without any prior knowledge. The deep Q Network \cite{mnih2015humanlevelcontrol}, Proximal Policy Optimization \cite{schulamn2017ppo}, and Advantage Actor-Critic \cite{mnih2016drpcitric} are well-known algorithms applied for learning the optimal control and operation strategies for power systems.
%The comprehensive review of the use of RL in power systems can be found in \cite{chen2021rddc}.
However, as the NN is integrated into the model, the DRL approaches inherit the vulnerability from NNs.

\textbf{Voltage regulation and topology control.} The DRL-based automation of power systems is threatened by cyber attacks and it might be overlooked by system operators. It motivates Chen \emph{et al.} \cite{chen2021safedrl} to discuss the vulnerability of DRL approaches applied in voltage regulation and topology control. They design a black-box, hard-to-detect data perturbation attack on sensor measurements. Based on the fact that the standard DRL framework does not take into account physical constraints like the power flow capacity, voltage magnitude limits, and power generation bounds, the attacker might make use of this fact to create adversarial observations. The observations are modified with small errors to bypass the detection of the error-checking mechanism and undermine the system's performance. A random distortion attack is proposed to maximize the optimal action before and after the attack and a target grid-aware attack is constructed to make the consequence follow a target system state trajectory. The  FGSM-like adversarial attack is constructed by querying the loss function to compute the gradient. However, the defect of this attack is that the control center can refuse to feedback the value of the loss function designed by the attacker.

\textbf{Topology optimization.} In \cite{zheng2021vldrlfirst}, it shows the small adversarial perturbation injected into the state variable can significantly affect the DRL-based topology optimization. The adversarial examples are generated with the conventional approach, i.e., FGSM, with cross-entropy loss function and bounded limitation. The criticality is derived from the impacts that are different when the attacks are executed at different moments. Therefore, instead of launching a continuous attack, the target periods are strategically chosen. Specifically, the time about when-to-attack is determined if the preferred action has a relatively larger (beyond a certain threshold) future return than all other actions.

\textbf{Emergency control.} In \cite{wan2022drllcps}, the DRL approach is used to solve the real-time emergency control problem for UVLS. The adversarial attack is constructed to mislead the recovery action (i.e., load shedding) after emergent events by perturbing the observed state within a certain threshold. The load percentage of the low load area is maliciously reduced to cause cascading failure or even widespread blackout or just do nothing after emergent events. As the voltage measurements are mutually restrained, the adversarial perturbation output from the FGSM or JSMA algorithm is forced to satisfy the physical constraints.

\textbf{Optimal power flow.} In \cite{zhou2020drlacopf}, it is reported that the DRL-based ACOPF is 7 times faster than the conventional solver considering a 200-bus test power system. However, in \cite{zeng2022pcvadrl}, it is shown that the DRL-based SCOPF solver has vulnerabilities in both the communication network and the model itself. An FDIAI attack strategy is designed to create adversarial observation states (continuous and discrete variables) against the DRL-based SCOPF. The attacker's goal is to minimize the return leading to ineffective control actions by adding small perturbations to the observed states. An optimization problem is formulated to minimize the adversarial perturbation with model misleading, BDD, and physical constraints, and it is solved by moving the misleading constraint to the objective function and throwing the other constraints to the objective function by multiplying them with penalty terms.
%Moreover, the authors prove that both the value-based and policy-based DRL approaches are vulnerable to adversarial examples.

\textbf{Demand response.} The DRL approach is also promising to be used in the demand response management system. In the demand response management system, each smart building often acts as an intelligent agent to flatten the demand curve and reduce the peak demand. A group of smart buildings can be treated as multiple intelligent agents to interact and achieve an overall supply-demand balance.  Therefore, the MARL can be used to address the challenges faced by inaccurate models and real-time decision-making requirements. To evaluate the vulnerability of the MARL-based demand response, the attacker is modeled as a single adversarial DRL agent \cite{zeng2022resiliendrl}. As the agents need the observations from the other agents to make a synchronous and joint decision, the malicious agent aims to learn an optimal policy to perturb his/her observation state to mislead the decision processes of other agents. Since the DRL-based application usually works in a feedback loop, the wrong operation or control strategy will affect the demand response system, which further increases the chaos among agents.
From the experimental result, the adversarial attack can cause a 41.43\% higher metric value of ramping than the no-attack case.

\subsection{Vulnerability of MLsgAPPs in the consumption scenario}\label{sec:consumption}
Next, we discuss the vulnerability of MLsgAPPs in the consumption scenario.

\subsubsection{Load forecasting}\label{section:lf}
Accurate prediction of load is of vital importance for both the financial benefit and system control in power systems. The increasing penetration of distributed energy resources also raises the urgent requirement for accurate load forecasting. However, the load variation is related to various factors such as temperature, weather, periodic effects, and government policies (e.g., COVID-19 restrictions). The optimal forecasting of loads seems to find a nonlinear and complex mapping between high-dimension factors and the time-series load. To address this issue, plenty of ML approaches have been used to achieve high performance of regression results for load forecasting \cite{kong2017dnnforecastload}. The complicated NN structures like CNN, LSTM, and GAN are increasingly used due to their significant capability of extracting hidden features from the historical data \cite{tan2019lstmloadforecas, wang2020ganloadforecast}.

As data integrity is critical for the regression task of load forecasting, the resulting load prediction might mislead the decision-making once the data integrity is breached. Luo \emph{et al.} \cite{luo2018lfbechrobust} construct a data integrity attack on the historical load data by multiplying the load with a Gaussian distribution error. It shows that the MAPE of the forecasted load is increased from 5.22\% to 10\% if 30\% of the load data are modified with a ratio of more than 40\%. However, the modification is so large that can be easily detected by system operators. To enhance the stealthiness of the attack, Niazazari and Livani \cite{remigijus2006lipschitz} design a data poisoning attack strategy to fool the outlier detector that is an NN-based regressor \cite{xie2016nndetector} and change the regression result obtained from MLR. The contaminated data is generated to maximize the prediction error of the test data. Since the parameters of MLR can be computed through the least square method, the sensitivity of the prediction against the input load is derived, which provides the direction for the modification of load data. The above method can be extended to the case where the FNN is used to forecast the load demand with the data of time and temperature. However, the construction of the adversarial attack depends on the model parameters and has poor transferability. Therefore, in \cite{zhou2022loadforeacting}, the adversarial attack is designed with gradient-based approximation or input-scaling approach, to maximize the squared loss or deviate the forecast to a specific direction \cite{zhou2022loadforeacting}.

Moreover, considering the case where the attacker has limited resources to execute the adversarial attack, Chen \emph{et al.} \cite{chen2019lfdja} propose a general framework to generate adversarial examples with limited knowledge of the target system. The adversarial attack is constructed by adopting the idea of PGD to either increase or decrease the forecast value in one direction as much as possible. Since the attacker does not know the target ML model, the gradient information is approximated based on the queried value, that is,
\begin{equation}\label{gradientapproximation}
  \nabla_{\emph{\textbf{x}}_k} f_{\bm{\theta}} \approx \frac{f_{\bm{\theta}}(\emph{\textbf{x}} + \xi \emph{\textbf{e}}_k) - f_{\bm{\theta}}(\emph{\textbf{x}} - \xi \emph{\textbf{e}}_k)}{2\xi},
\end{equation}
where $\emph{\textbf{x}}_k$ is the $k^{th}$ element of $\emph{\textbf{x}}$, $\emph{\textbf{e}}_k$ is a vector with 1 at the $k^{th}$ position and 0 elsewhere, and $\xi$ is a scaling factor. It is proved that the load forecasting performance degrades drastically by only perturbing very few elements of $\emph{\textbf{x}}$. Besides, the slightly perturbed and few-degree target perturbation of temperature data could result in increased operating costs and load shedding. A greedy search algorithm is proposed to find the most vulnerable loads and compute the corresponding adversarial perturbations to increase the resulting operation cost. Using a similar idea, Chen \emph{et al.} \cite{chen2022fistml} evaluates the vulnerability of the RNN-based building load forecasting with a modified FGSM algorithm by making the untarget components of input zero. However, the performance of adversarial attacks depends on the query times to estimate the gradients, which might cause a delay in launching the attack and be detected due to the frequent queries.

\subsubsection{Energy theft detection}\label{section:etd}
Energy theft is one of the critical threats that can cause non-technical financial losses for power utilities. To defend against energy thefts, advanced smart meters are deployed to monitor the consumer's electricity consumption. By taking advantage of the massive fine-grained power consumption data, the DL approaches are applied to achieve state-of-the-art energy theft detection performance. However, the information and communication facility opens a door for attackers to take over the smart meter \cite{zanetti2017tunableami}. They can penetrate the smart meter to modify the power consumption data \cite{awareness2018hack,sun2017smartmeterhack} and the DL-based energy theft detection can be cheated by the adversarial examples \cite{li2020dletd, li2020searchfromfree}.

Except for evasion attacks, the ML-based energy theft detector is also vulnerable to the data poisoning attack. A comprehensive study is conducted in \cite{takiddin2021retd} to compare the weaknesses of different ML models. It is validated that the unsupervised learning approaches such as the ARIMA, autoencoder with attention, and supervised learning approaches such as the random forest, Adaboost, SVM, and FFNN, are all vulnerable to data poisoning attacks. However, the DL approaches perform much better than the shallow learning approaches under adversarial attacks.

\subsubsection{Non-intrusive load monitoring}\label{section:nonintrusiveloadmonitoring}
As a huge amount of electricity usage data is collected, the power consumption (e.g., real power, current, voltage, and energy) of a dwelling can be monitored non-intrusively. Hence, the active electric appliances can be inferred without deploying dedicated sensors. Generally, the NILM outputs a metric to quantify the flexibility of consumers as participators in the demand response. Using the data of NILM, the power usage patterns can be inferred using the NN-based approaches (e.g., LSTM, CNN, and RNN), which will benefit the overall situational awareness and high-quality response to the users' demands.

However, Wang and Srikantha \cite{wang2021blackboxnilm} recognize the deficiency of ML models and propose an adversarial attack to make the predicted active appliances different from the real active ones by inserting small perturbations into the power consumption data. Considering a black-box scenario, an FFNN-based substitute model is trained using the universal approximation theorem to make up for the shortage of not knowing the target model. The true inference of active appliances is queried and then used to expand the dataset and calculate the value of the loss function for the substitute model. The data augmentation process is conducted by iteratively updating the data point with the momentum information, which is given by:
\begin{equation}\label{eq:dataaugmentation}
  \mathcal{S}_{\rho+1} \leftarrow \{ \emph{\textbf{x}}^{k+1} = \emph{\textbf{x}}^{k} - \emph{\textbf{v}}^k, \emph{\textbf{v}}^{k+1} = \alpha\emph{\textbf{v}}^k + \lambda \nabla_\emph{\textbf{x}}(f_{\bm{\theta}}(\emph{\textbf{x}}, y)) \}\cap \mathcal{S}_{\rho},
\end{equation}
where $\mathcal{S}_{\rho+1}$ is the augmented training dataset, $\rho$ and $k$ are iteration numbers, $\emph{\textbf{v}}$ is the momentum variable, and $\lambda$ is a tunable parameter. The adversarial perturbation is constructed iteratively and its magnitude is constrained by a certain threshold. Since the incorrect prediction drives the imbalance in demand and supply, the power system would lose its order in control and operation under the adversarial attack.

\begin{figure*}[htbp!]
\begin{center}
\includegraphics[width=0.7\textwidth]{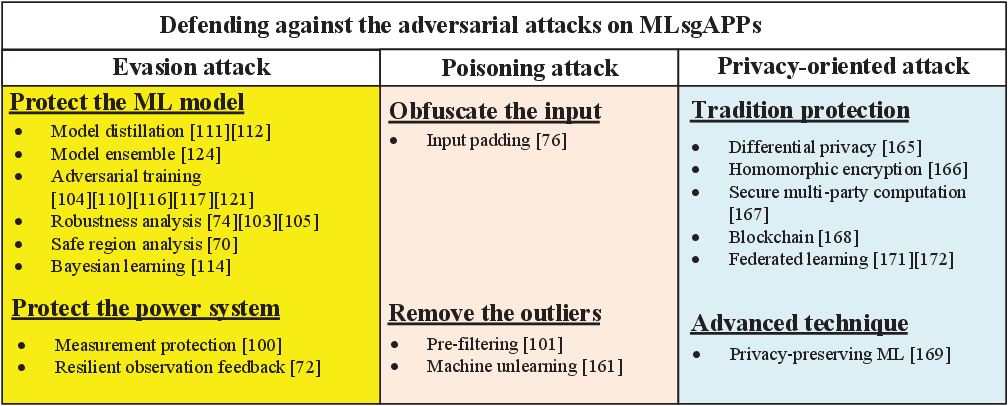}
\caption{A summary of defense approaches for protecting MLsgAPPs.}\label{fig:summarizethedefenseapproache}
\end{center}
\vspace{-0.5cm}
\end{figure*}

\subsection{Some takeaways about the adversarial attacks}\label{takeawayssome}
The researches about the adversarial attacks on MLsgAPPs evolves in three stages. In the first stage, it is recognized that the MLsgAPPs suffer from adversarial attacks. The general attack strategies are introduced to evade and poison the MLsgAPP model. In the second stage, the specifics of power systems are considered as constraints for designing the adversarial attacks. The adversarial attack can only be successful if the attacker considers factors such as the system dynamics, bad data detection, execution environment, and physics laws. In the third stage, the attack's impact on the operation and control of power systems is considered an objective for designing adversarial attacks. The adversarial attacks on MLsgAPPs are practical only if they can cause adverse effects on the operations such as frequency/voltage control, generation cost, and carbon emission.

\section{Defense Approach}\label{section:defensiveapp}
Regarding the threat introduced by adversarial attacks, there are countermeasures developed to protect MLsgAPPs. In the CV domain, there are many state-of-the-art defense approaches such as adversarial training \cite{kurakin2016amlatscale}, model distillation \cite{papernot2016distillation}, adversarial detection \cite{ma2019nic}, and input reconstruction \cite{meng2017magnet}. In \cite{machado2021amlic}, the related studies are classified from 6 aspects including gradient masking, auxiliary detection, statistical comparison, preprocessing, an ensemble of classifiers, and proximity-based detection. These defense approaches can be naturally extended to protect MLsgAPPs. Since the ML models suffer from adversarial attacks inevitably, it is of vital importance to enhance and improve the security of MLsgAPPs. We summarize the defense approaches in Fig. \ref{fig:summarizethedefenseapproache}. The defense approaches against evasion attacks are reviewed from the aspects of protecting the ML model and power data. There are two approaches to defense against the poisoning attacks, which are the obfuscation of the input and the removal of outliers. For privacy-oriented attacks, the defense methods adopt the traditional techniques (e.g., differential privacy, homomorphic encryption, and blockchain) and the recent privacy-preserving NN for protecting the input data and model parameters.

\subsection{Defending against evasion attacks}

We first review the countermeasures against evasion attacks.

\textbf{Model distillation.} The robustness of the NN-based energy theft detection model is enhanced by using the predictions as labels to train a distilled model \cite{li2020dletd, li2020searchfromfree}.

\textbf{Model ensemble.} A sequential combination of an autoencoder with LSTM cells, recurrent layers with GRUs, a fully connected layer, and an output layer is proposed to improve the detector's sensitivity against the perturbed power consumption data \cite{takiddin2021retd}. However, although the combined model is complicated, the performance of energy theft detection has deteriorated slightly by 1-3\%.

\textbf{Adversarial training.} Among the defense approaches, adversarial training is the most famous approach used to improve the robustness of MLsgAPPs \cite{niazazari2020attgec, tian2022anatt, tian2022aacnnpqr}. Ren \emph{et al.} \cite{ren2022vulpsstvs} consider the black-box adversarial training to train an ensemble surrogate ML model to mitigate adversarial examples. However, this work assumes an impractical case where the defender does not have any knowledge of the implemented ML model of its own. That is, the defender and ML service are different parties or belong to different departments of a power utility.

To enhance the resilience of the MARL-based demand response, Zeng \emph{et al.} \cite{zeng2022resiliendrl} propose a periodic robust adversarial training approach to learn the experience for dealing with adversarial attacks. Specifically, the adversarial training is conducted based on a robust Markov game through min-max optimization \cite{pinto2017rarl, zhang2021rrl}. This is a novel idea as adversarial training is first used to enhance the robustness of the MARL-based demand response management system. By treating one agent as the adversary, the resiliency enhancement approach consists of three parts: i) training the optimal adversary policy for the adversarial agent; ii) training the overall policy with a fixed adversary's policy; iii) conducting the adversarial training periodically. The adversary's policy is to minimize the overall reward computed according to the district demand balance with an opposite reward value compared to the other agents and a disturbing observation. However, the adversary is assumed to know all parameters of the MARL model, system model, and defense strategy, which might not be practical in the scenario with real-time and fast interactions.

\textbf{Robustness analysis.} As most robustness-enhancement approaches are specially designed for specific adversarial attacks, a general lower bound of robustness is derived independent of adversarial algorithms \cite{Szegedy2013rss,katz2017reluplex}. To find an attack-independent robustness index for MLsgAPPs, Ren and Xu \cite{ren2022vulverifications} verify the adversarial robustness by deriving the lower bound of the adversarial perturbation of a specific instance with the assumption of local Lipschitz continuity. The non-differentiable constraint is resolved using the backward pass differentiable approximation when the NN model contains ReLU units. However, it is difficult to compute and achieve the lower bounds, because the bounds are related to the datasets, parameters of the ML model, and parameters of the power system model.

Besides, the above robustness analysis does not consider the impact of specifics in power systems. The adversarial examples must follow the physics laws and the error-checking mechanisms for filtering bad data caused by either system faults or cyberattacks. Therefore, in \cite{zhang2022pcrvisaps}, a systematic framework is proposed to analyze the impact of physical constraints and BDD on the robustness of the ML-based security assessment model. It is observed that the ML models become more robust against adversarial examples as the input cannot be arbitrarily perturbed, and the attacker's limited knowledge of power system and sensor measurements can further reduce his/her capability of successfully launching the adversarial attacks \cite{zhang2022corestability}. The results show that once the system operator can protect a few vulnerable sensor measurements, the model's robustness can be improved largely. However, the obtained conclusions are restricted to the specific scenario and dataset. A general derivation is missing, which makes it hard to precisely locate the robust areas for the input of MLsgAPPs.

\textbf{Safe region analysis.} Normally, the lower bound of adversarial perturbations is evaluated and verified from the attacker's perspective but standing on the defender's side. Venzke and Chatzivasileiadis \cite{venke2021verifynn} examine the input regions where no adversarial example exists. The maximum safe region of the ML-based security assessment model is computed by transferring the non-differentiable constraint caused by ReLU units to a MILP problem. To reduce the computational complexity, the authors introduced the interval arithmetic \cite{Tjeng2019ia} and weight matrix sparse approaches \cite{zhu2018sparsematrix}. However, the numerical results are also specified by the test dataset, which might be far from the real data distribution.

\textbf{Bayesian learning.} In \cite{zhou2022loadforeacting}, Bayesian learning is adopted for its property to vanish the gradient-based attacks. The authors provide analytical proof that the Bayesian NN is more robust than the traditional NN model considering the regression task for load forecasting. The mathematical results are derived based on an ideal assumption of infinite training data and model parameters, which show that the model's sensitivity against the input is a key factor for that the minor perturbation can make the prediction deviate a lot from the expectation. In our opinion, the Bayesian learning model treats all parameters in an NN as random variables. Therefore, it outputs a distribution rather than a fixed value, which might reduce the possibility for the attacker to successfully construct adversarial examples against the model.

\textbf{Measurement protection.} From the angle of traditional protection against cyberattacks on power systems, Tian \emph{et al.} \cite{tian2021seaann} propose protecting sensor readings using the technologies of encryption, authentication, and access control. It is derived that different sensors have different importance rankings in terms of preventing attacks. However, the derivation of the importance ranking of sensors depends on a differentiable NN model, which reduces the scalability of this approach to the other ML model structures.

\textbf{Resilient observation feedback.} To produce reliable control actions for the DRL-based UVLS under adversarial attacks, Wan \emph{et al.} \cite{wan2022drllcps} propose a distance-based resilient observation feedback method using the historical states instead of the current state. With the assumption that the attacker usually modifies the latest observation, the furthest observation that is furthest from the current one might be the true observation . However, this defense method might affect the convergence of the DRL algorithm and the accuracy of control actions.

\subsection{Defending against poisoning attacks}
The countermeasures against evasion attacks have been widely studied, but the defense against poisoning attacks has limited investigation. Here we review the countermeasures against data poisoning attacks.

\textbf{Input padding.} In \cite{li2021adnn}, the adversarial attacks are detected by randomly padding the model input in both the training and prediction stages.  The input is expanded with zeros randomly placed at positions in the column or row of a batch. To maintain the physical properties of normal measurements, the zeros are padded before or after the measurement vector. Since the iterative attack has poor transferability, the adversarial attack will fail if there are changes in the model input. The input padding works because it is difficult for the attacker to obtain complete information about the model input. However, since the padding is executed on all inputs, it might not be able to amplify the abnormality of the adversarial example.

\textbf{Pre-filtering approach.}  To filter out contaminated training data, Kamal \emph{et al.} \cite{Kammal2021ppaclsevent} propose an unsupervised detector to detect outliers caused by poisoning attacks. The numerical results prove that the linear kernel of the SVM classifier is more robust than the nonlinear kernel in the face of a poisoning attack on the SVM-based grid event classification \cite{Kammal2021ppaclsevent}, although the linear kernel has poorer classification accuracy. With the assumption that the training data is uniformly distributed, the larger the difference in the classification boundary between the classifier and the oracle classifier, the more suspicious that the corresponding subset of the training set has contaminated data. However, the above approach only focuses on simple ML models, which might fail to distinguish the poisoned data from normal data if the scenario is complex.

\textbf{Machine unlearning.} Recently, machine unlearning has been a very popular research direction \cite{xu2023machine}. The basic idea of machine unlearning is to retrain the ML model with a subset of the training set and try to reduce the impact of the removed training samples. Considering the data poisoning attack, the poisoned samples can be removed by selecting a partial of the training dataset to retrain the ML model. Alternatively, the poisoned samples are given smaller weights contributing to updating the model \cite{xu2023task}. This is an effective approach if we know the bad or malicious data in the training dataset.

\subsection{Defending against privacy-oriented attacks}
Although the privacy issue of power data has been considered for decades \cite{asghar2017smart}, the protection of sensitive/private information of MLsgAPPs is seldom considered. There  are studies devoted to the inference of private information in the training data \cite{shokri2017meminfer}, input \cite{wood2020homomorphic}, and model parameters \cite{gong2020model} in the ML field. Traditionally, the privacy issue of power systems is mainly about protecting the collected data for power system applications. Data privacy mainly refers to the consumers' and/or utilities' private and secret information. The widely studied privacy-preserving approaches include differential-privacy \cite{fioretto2019differential}, homomorphic encryption \cite{wu2021privacy}, secure multi-party computation \cite{lu2012eppa}, blockchain \cite{guan2018privacy}, etc. A pioneer work on the data privacy of MLsgAPPs adopts the homomorphic-encryption-friendly NN to conduct the residential electrical load forecasting for protecting the user data and model parameters \cite{wu2023sectcn}. However, the approximated ReLu function makes the ML model not accurate and causes an MAPE larger than 10\%. The reduction of efficiency also affects the real-world use of this kind of privacy-preserving MLsgAPPs.

Besides, the federated learning is a promising learning approach to protect the privacy of users' local information \cite{9084352}. The large-scale power grid usually has multiple centers, distributed substations, and different operation utilities, which have plenty of data that is either secret or private. Therefore, federated learning has been widely adopted to address the issues caused by data islands and privacy in power systems \cite{zhang2022federated}. Especially on the consumer's side, the data protection laws strictly require that the collection of users' data must follow the privacy-preserving rules. Hence, there are a lot of papers about federated learning used in the scenario of consumption \cite{9729772}.

\subsection{Some takeaways about the defense approaches}\label{takeaways}
The defense is conducted from the perspective of either the ML model or the power system. Specifically, the approaches used to defend against the adversarial attacks on ML models can be naturally applied to defend against the adversarial attacks on MLsgAPPs. The specifics of power systems are considered for analyzing the robustness and safe region of MLsgAPPs under attack. On the other hand, we can prevent the input of MLsgAPPs from being modified by enhancing defense strategies such as measurement protection \cite{deng2015defending} and moving target defense \cite{liu2021converter}\cite{zhang2022security}. Using the Trust zone \cite{pinto2019demystifying} or secure historian \cite{mcnamee2011secure} can also defend against adversarial attacks. Encryption \cite{zhang2020secure} and authentication \cite{yang2023secure} are easy-to-deployment alternatives although they might affect the real-time requirement for the power system operation. As such, the goal of most defense approaches focuses on protecting the input and model parameters of MLsgAPPs. The countermeasures designed by combining the traditional defense approaches for ML models and power system applications might avoid the shortcomings.

\section{The Dataset Of MLsgAPPs}\label{section:datasetpowersystem}
The dataset used by MLsgAPP is of significant importance for training a good model and plays an important role in analyzing the vulnerability of MLsgAPP. In the following, we classify the applied dataset into two types: the real-world dataset and the synthetic dataset.

\subsection{Real-world dataset}
The real-world dataset is generated by systems running in reality. The load data collected by the 2012 Global Energy Forecasting Competition\footnote{https://www.kaggle.com/c/global-energy-forecasting-competition-2012-
load-forecasting/data.} is used to create voltage measurements and active/reactive power flows \cite{tian2021seaann}, which is directly used by \cite{remigijus2006lipschitz} to evaluate the vulnerability of NN-based load forecasting. The micro-PMU measurements from a feeder in Riverside, CA, are applied for the SVM-based grid event classification \cite{Kammal2021ppaclsevent}. The historical load and temperature data from the ISO New England are used by \cite{zhou2022loadforeacting} to validate that the Bayesian NN model is more robust than the other NN models. To test the adversarial performance \cite{chen2019lfdja}, the actual load dataset is collected from European Network of Transmission System Operators for Electricity API\footnote{https://transparency.entsoe.eu/} and the weather data are queried from Dark Sky API\footnote{https://darksky.net/forecast/47.3769,8.5414/us12/en}. The data published by the Irish Social Science Data Archive is used as the benchmark dataset for testing the energy theft detectors \cite{li2020dletd,li2020searchfromfree}. The dataset of Almanac of Minutely Power\footnote{http://ampds.org/} and Pecan Street\footnote{https://dataport.pecanstreet.org/} are used to learn the NILM model \cite{wang2021blackboxnilm}. The public data provided by Elexon's BM reports\footnote{https://www:bmreports:com/bmrs/?q=help/about-us[Accessed10/07/2020]}, ENTSO-E's data transparency platform\footnote{https://transparency:entsoe:eu/}, National (British) Grid's data explorer\footnote{https://data:nationalgrideso:com/} and the Nordpool website\footnote{https://www:nordpoolgroup:com/} are combined to generate the required dataset for the inertia forecasting \cite{chen2022costoriented}. The load data gathered from NYISO is a well-known dataset that can be cooperated with the IEEE standard test power system to generate the dataset for designing FDIAs \cite{sayghe2020adlpsse}.  The Irish Smart Energy Trail dataset published by the Sustainable Energy Authority of Ireland contains 3000 users' smart meters that took consumption data every 30 minutes over 18 months\footnote{Irish Social Science Data Archive, Available: http://www.ucd.ie/issda/data/
commissionforenergyregulationcer/}, which is treated as the base data to generate energy theft data\footnote{``Pecan street inc. dataport 2019,'' Website: https://www.pecanstreet.org/dataport/papers/, 2023.}.

\subsection{Synthetic dataset}
The synthetic dataset is generated by combining the data originating from simulators and the data created by physical models. The DSATools can simulate stable and unstable operating points for evaluating the dynamic security combined with the IEEE standard power systems \cite{ren2022vulpsstvs,ren2022vulverifications}. The MATPOWER provides test cases of power systems with data collected from real-world power facilities (e.g., the England and Polish standards) \cite{zimmerman2010matpower}. It integrates well-tested functions for calculating the DC and AC power flows. Therefore, MATPOWER has been widely used to generate high-fidelity power data \cite{venke2021verifynn}. For example, the ground-truth voltages are estimated through the AC power flow equations solved by MATPOWER \cite{tian2021seaann} and the FDIAs are constructed based on the test cases \cite{li2021conaml, liu2019annsg, li2021adnn, sayghe2020adlpsse, tian2021jae, tian2021amlfdia}. The real-time digital simulator (RTDS) is another famous simulator that can be used to generate transient power data \cite{niazazari2020attgec}. The MATLAB PSTv3.0 can be used to create datasets for the task of data-driven CCT prediction \cite{liu2021gandiam}. The power quality data is generated according to the physical models for the sag, impulse, and distortion signals \cite{tian2022anatt,tian2022aacnnpqr,igual2018pqdis}.

Besides, there are many open-source platforms to investigate the feasibility of implementing state-of-the-art ML approaches to operate a power system autonomously. The L2RPN\footnote{https://l2rpn.chalearn.org/} challenges open opportunities for researchers all over the world. It provides an environment for competitors to run ML algorithms to control the power system \cite{marot2020L2prn}. Up to now, the competitions have been launched in IJCNN 2019, WCCI 2020, NeurIPS 2020, ICAPS 2021, and WCCI 2022. The involved datasets have been used to evaluate the vulnerability of the applied DRL approaches \cite{zeng2022pcvadrl,zheng2021vldrlfirst}. The OpenAI Gym environment CityLearn provides a convenient platform for researchers to evaluate their MARL algorithms for power system applications \cite{zeng2022resiliendrl,nagy2021citylearn}. Gym-ANM is another well-known simulator for emulating the power system control with ML approaches \cite{henry2021gymanm}. The RLGC is the world's first open-source platform for testing DRL approaches for the operation and control of power systems \cite{wan2022drllcps,huang2019rlgc}. Moreover, the building load consumption data is generated according to the building simulation platforms using the EnergyPlus \cite{chen2022fistml}.

\section{Future research directions}\label{section:futuredirection}
ML approaches have been proven academically efficient for power system operations. Yet it is still not clear how to practically apply the learning approaches to real-world devices or processes. The unavailable, absent, or publicly accessed operation data and codes of power systems impede the safe use of ML approaches. Therefore, the vulnerability analysis of MLsgAPPs is still an open problem to be considered. Here we discuss some future research directions from the aspects of the attack and defense.

\subsection{Adversarial attack}\label{section:adversarialattack}
The following issues should be further discussed for constructing adversarial attacks on MLsgAPPs.

\emph{\textbf{Feasibility.}}
The feasibility of launching adversarial attacks on MLsgAPPs is the first issue. The natural limitations raised in power systems set tough barriers for attackers to cross through. The design of adversarial attacks is different from the other scenarios where the input is image or text. For example, the input can be very complicated in power systems, which is even confidential information considering the commercial secret. Besides, the complex power system has many nonlinear relationships between power variables. The lack of mathematical derivation makes the vulnerability assessment of MLsgAPPs highly dependent on the dataset, which lacks generality since the analysis is case-by-case.

\textbf{\emph{Attack model.}}
From the perspective of the ML model, the existing studies mainly focus on the model-oriented attacks on MLsgAPPs. As we have mentioned in Section \ref{section:vulnerableML}, there are also privacy-oriented and platform-oriented adversarial attacks. In power systems, the applications in the home-area energy management system involve plenty of private and secret data. These data are either highly-valued commercial data or users' sensitive information. Once they are used for training the MLsgAPPs, there are severe privacy issues. The fine-grained membership inference method can expose the economic condition of the power utility and the user's preferences.
%Moreover, the secret management data and the factory's operating data of the power corporation might be obtained by spies or adversaries from opponents.
Moreover, the running platform of MLsgAPPs is composed of general computers and servers, which means that zero-day vulnerabilities or old and unpatched vulnerabilities are big threats. Besides, the implementation software of ML algorithms sometimes is crawled from a third-party open-source website, which might be embedded with backdoors or have bugs in the codes by careless engineers. Therefore, the codes should be carefully audited and checked before being used for MLsgAPPs.

From the perspective of power systems, the reviewed papers usually have the assumption that the attacker has real-time and full knowledge of the power system. However, the assumption is so strong that might not be practical. For example, most adversarial attacks are constructed with specific system structures and parameters, which might change at a high frequency for the operation and control of the grid.  Therefore, it is still an open problem to design adversarial attacks on MLsgAPPs when the attacker has incomplete and even zero knowledge of the power system.

%Besides, some studies have discussed the case where the attacker has limited capability/resources to modify the model input. This is also a more reasonable assumption for attackers as the power system is so complicated and heterogeneous that the attacker cannot compromise every device.

\textbf{\emph{Attack impact.}}
The analysis of the impact of the adversarial attack on MLsgAPP on the operation of the power system is an important issue. In \cite{wang2021blackboxnilm}, it has been proved that the resource-constraint attacker can cause infeasible operating conditions, loss of equipment, and large-scale load-shedding if the DRL-based operation and control are affected by adversarial examples. However, the self-healing capability of the power system can defend against some adversarial attacks naturally. Considering the adversarial attack on the DL-based energy theft detection, the studies \cite{li2020dletd} and \cite{li2020searchfromfree} show that it is the stolen profit that matters to construct adversarial examples rather than the divergence between the original and crafted measurements.

Therefore, the adversarial attacks on MLsgAPPs are effective only if they cause explicit impacts such as economic loss, load shedding, generation dispatch, line overloading, nodal price change, etc. If the adversarial attacks do not cause any impact on the system's operation, they fail to achieve the goal of damaging the system. In \cite{chen2022costoriented}, the operation cost is considered as the consequence caused by the adversarial attack on the ML-based inertial forecasting. Thus, the attacking goal is not to maximize the forecasting error but to approach the target operation cost. This is a reality for constructing adversarial attacks on MLsgAPPs.

%Besides, the robustness of MLsgAPPs should also include the impact of the adversarial attack on the system operation. For example, the operation cost caused by the adversarial attack. In that case, it is more reasonable to be used to quantify the robustness of AIpsAPPs.
\textbf{\emph{Timely attack.}}
The attack timing is a critical issue in launching adversarial attacks on MLsgAPP. First, it is important to guarantee that the adversarial attack is finished timely such that the attack takes effect at the right period. The power system runs in real-time within a few minutes, seconds, or milliseconds of time scale. If it is computationally intensive to calculate the adversarial example, the attacker might miss the correct period and be identified by the time-evolution detectors. Second, the appropriate time to launch the adversarial attack is also important. If the timing is well selected, the attack can cause the maximum damage and reduce the continuous effort required to achieve the objective. For example, considering the dynamic interaction processes of the DRL algorithm, the timing for launching the attack is very important. Once the power system is attacked at the most sensitive moment, it might face hazardous situations.

\textbf{\emph{Attack on the implementation platform.}} Moreover, the security of the implementation platform of ML algorithms sometimes is ignored. The information infrastructures supporting the processing, training, and predicting of MLsgAPPs are also vulnerable. The malicious insiders can change the model parameters, steal the model information, and place backdoors in the hardware and software of the implementation platform. A direct attack route is just to hack into the computers or servers in the computing center to take over the execution of MLsgAPPs. Therefore, the defense against cyberattacks on the implementation platform of MLsgAPPs is also important and cannot be ignored.

\subsection{Defense approach}\label{section:defenseapproach}
Although the adversarial attacks on MLsgAPPs have been widely studied, the corresponding defense approaches have not been much investigated. In this section, we discuss the future research direction to enhance the resiliency of MLsgAPPs.

\textbf{\emph{Resilient ML.}}
Although adversarial training is effective in defending against known attacks, it comes at a cost of accuracy degradation on clean data. The model might be overfitted after the retraining with adversarial examples.
%The performance of the resiliency improvement also depends on the generated adversarial examples feedback to the training process.
Considering the power system context, the power data changes a lot due to the periodic variation of power demands (e.g., the peak demand each day and the seasonal factors) and the environment (e.g., the weather, sunlight, and wind speed). Since the prediction/classification boundary changes, the adversarial examples defined in this period might be normal data in the next period. The retraining processes should be frequently conducted which might add computation burden on the system's operation and control. Therefore, the adversarial training process should be power context-aware, that is, the adversarial examples have good transferability or can be easily updated according to the variation of the power system's behavior.

Besides, adversarial detection is an effective approach to finding outliers in the training dataset and model input. The unsupervised ML model can be trained to filter out abnormal data. It seems that the approach is effective because the abnormal data and normal data have different distributions. However, in power systems, the malicious sensor measurements can be carefully designed to lie in the cluster of normal data \cite{liu2020dummyatt,mohamd2020ganfdia}. Therefore, the adversarial examples might hide in the training dataset or bypass the detection. Recently, the physics-informed neural network has been widely adopted to extract the dynamic features of power systems \cite{huang2022pinn}, which might provide a potential idea to develop adversarial detection approaches for MLsgAPPs.

%From the above review, the robustness of MLsgAPPs is analyzed with a given dataset and specific assumptions. A general analytical conclusion is missing, which makes it hard to precisely locate the robust areas for the input of MLsgAPPs. Besides, since the robustness evaluation problem considering the system constraints is usually nonconvex, the approximated results might affect the accuracy for the determination of which model is more suitable for a specific task.

%Considering the privacy issue, the federated learning (FL) has been used in different power system applications \cite{su2022flcola}\cite{wen2022feddetect}. However, these studies mainly focus on the privacy of the users' power consumption data. The privacy of the model input (e.g., the bidding price of each renewable energy resource), model structure, and model parameters are overlooked. The secrecy of the ML model is important for the commercial use, which can be protected using the cryptography technologies.

\emph{\textbf{Resilient power system.}}
Apart from improving the resiliency of ML models, the security enhancement of the power system itself is also of vital importance. Since it is impossible to close all entry points against attackers, a potential idea is to increase the difficulty or cost of launching the attack. The traditional approaches such as encryption, authentication, and randomization, and the defense approaches for ML approaches should be coordinated. For example, the measurement padding and moving target defense-based approaches \cite{zhang2019analysis} can be combined to reduce the possibility for the attacker to successfully launch adversarial attacks.

The robust control-based preventive method can treat maliciously generated biases as unknown uncertainties and the robust controller is designed to ensure that the tracking error under attacks could be bounded, which
typically requires no other investments besides inducing some extra computation burdens \cite{zhou1998essentials}. Sadbadai \textit{et al.}  \cite{9600614} designed a series of distributed cyber-resilient controllers for (parallel) DC and AC microgrids (focusing on frequency regulation and active power sharing) to mitigate the adverse impact resulting from the bounded FDIAs against secondary communication links and actuator signals. The key challenge of designing robust control strategies to defend against adversarial attacks is the theoretical analysis of control performance in the presence of intractable adversarial injections.

\subsection{Vulnerability of large language model-based smart grid applications}\label{section:llmsgapp}
Recently, the large language model (LLM) has created a great wave to apply the complicated and large NN models in different areas. There is a pioneer work \cite{bonadia2023potential} that demonstrated that the ChatGPT can create distribution systems for simple load flow studies. More recently, Li \emph{et al.} \cite{li2023framework} evaluates the capability of ChatGPT to create codes for solving the unit comment problem and decentralized optimization. The studies show fantastic results about the use of LLM to help educators and engineers. However, it is revealed that the LLM is facing with the threat of prompt injections, remote code execution on the backend system, poisoning training data, denial of service, supply-chain backdoors, jailbreaking attacks, etc.\footnote{https://www.tarlogic.com/blog/owasp-top-10-vulnerabilities-llm-applications/
} The vulnerability of LLM has also attracted attention from researchers \cite{shi2023badgpt,gupta2023chatgpt}. One recent study is about the jailbreaking attack on the LLM chatbots, where malicious users manipulate the prompts to uncover sensitive, proprietary, and even harmful information against the usage policies \cite{deng2023jailbreaker}. The vulnerability of LLM indicates that its adversarial issues must be taken into consideration once it is used in power systems.

\section{Conclusion}\label{section:conclusion}
In this paper, we conducted a comprehensive review of studies about the vulnerability of ML-based smart grid applications (MLsgAPPs). Taking the adversarial attacks seriously, we first validated that the attacks on ML models and power data were real threats using facts from academia and industrial investigations. The specifics for evaluating the security issues of MLsgAPPs were highlighted. Then, the existing studies about the adversarial attacks on MLsgAPPs were reviewed in detail and the similarities and differences were summarized from different aspects. We also surveyed the literature about the countermeasures to defend against adversarial attacks on MLsgAPPs. Moreover, we collected the datasets applied in MLsgAPPs for interested followers. Finally, we provided future research directions in further assessing the vulnerability of MLsgAPPs.

\begin{spacing}{1}
\footnotesize
\bibliographystyle{IEEEtran}
\bibliography{bibliography}
\ifCLASSOPTIONcaptionsoff
  \newpage
\fi
\end{spacing}

\end{document}